\documentclass[lettersize,journal]{IEEEtran}

\usepackage{amsmath,amsfonts}
\usepackage{cite}
\usepackage{algorithmic}
\usepackage{array}
\usepackage[caption=false,font=normalsize,labelfont=sf,textfont=sf]{subfig}
\usepackage{enumitem}
\usepackage{textcomp}
\usepackage{stfloats}
\usepackage{url}
\usepackage{verbatim}
\usepackage{graphicx}

\def\BibTeX{{\rm B\kern-.05em{\sc i\kern-.025em b}\kern-.08em
    T\kern-.1667em\lower.7ex\hbox{E}\kern-.125emX}}

\usepackage{amsmath, amssymb}
\usepackage{amsfonts}
\usepackage{tikz}
\usepackage{multirow}
\usetikzlibrary{arrows.meta, positioning, calc}
\usepackage{xcolor}
\usepackage{booktabs}
\usepackage{tabularx}
\usepackage[
    colorlinks=true,
    linkcolor=blue,
    citecolor=blue,
    urlcolor=blue
]{hyperref}
\usepackage{xr-hyper}
\usepackage{cleveref}

\newif\ifincludesupplement
\includesupplementfalse 

\crefname{section}{Section}{Sections}
\crefname{figure}{Fig.}{Figs.}
\crefname{table}{Table}{Tables}
\crefname{equation}{Eq.}{Eqs.}

\definecolor{inputgray}{RGB}{228, 228, 228}
\definecolor{ttkblue}{RGB}{188, 216, 238}
\definecolor{omegagreen}{RGB}{192, 226, 192}
\definecolor{measurepurple}{RGB}{216, 198, 233}
\definecolor{solverorange}{RGB}{253, 223, 183}
\definecolor{piblue}{RGB}{193, 213, 243}
\definecolor{xiorange}{RGB}{253, 208, 162}
\definecolor{distred}{RGB}{240, 193, 193}
\definecolor{xiaccent}{RGB}{190, 60, 60}
\definecolor{arrowcolor}{RGB}{100, 100, 100}

\newcommand{\para}[1]{\vspace{1mm}\noindent{\textbf{#1}}}

\title{MS-COOT: Comparing Morse-Smale Complexes with Co-Optimal Transport}

\author{Guangyu Meng$^{*}$, Mingzhe Li$^{*}$, and Erin Wolf Chambers%
\thanks{Guangyu Meng, Mingzhe Li, and Erin Wolf Chambers are with the University of Notre Dame (e-mail: \{gmeng,mli33,echambe2\}@nd.edu).}%
\thanks{$^{*}$Guangyu Meng and Mingzhe Li contributed equally to this work.}%
}

\graphicspath{{figs/}{figures/}{pictures/}{images/}{./}} 

\begin{document}

\maketitle

\begin{abstract}
  Understanding and comparing structures in scalar fields is a central challenge in scientific visualization, with applications ranging from feature analysis to temporal and structural comparison. The Morse-Smale (MS) complex provides a natural representation by decomposing a scalar field into regions induced by gradient flow. 
However, existing approaches typically rely on graph-based representations, matching critical points without explicitly establishing correspondences between regions. 
In this work, we represent the MS complex as a hypergraph, where critical points form nodes and regions define hyperedges. 
We introduce MS-COOT, a co-optimal transport distance that jointly computes correspondences between critical points and regions. 
This formulation enables explicit region-to-region matching within a distance-based framework, allowing identification of region-level events such as splitting and merging. 
We instantiate this framework with domain-specific components, including a hypernetwork function encoding relationships between critical points and regions, persistence-based probability measures, and a sample cost term that incorporates critical point attributes. 
We evaluate MS-COOT on datasets spanning 2D simulations, 3D surface meshes, and volumetric data. 
Our results show that MS-COOT provides distinct characterization of certain region-level structural changes compared to graph-based distances, achieves competitive accuracy in shape classification, and improves classification of simulation parameter settings in a 3D ensemble.

\end{abstract}


\begin{figure*}[!t]
    \centering
    \vspace{-2mm}
    \includegraphics[width=0.98\textwidth]{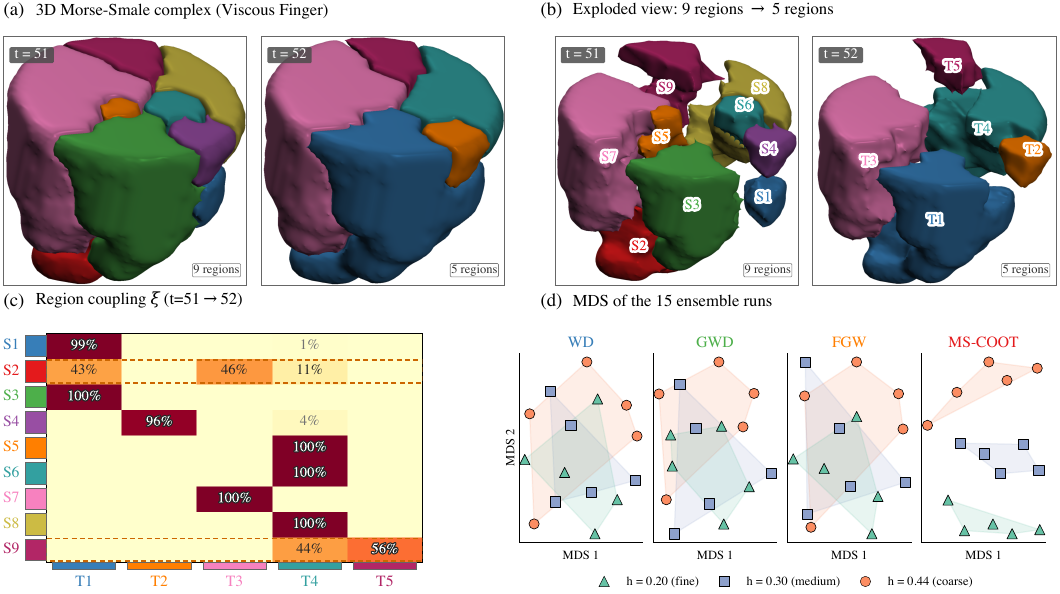}
    \vspace{-2mm}
    \caption{\textbf{MS-COOT produces region-to-region correspondence between 3D Morse-Smale complexes.} (a)~3D Morse-Smale complexes extracted by TTK~\cite{tierny2017topology} from a Viscous Finger simulation~\cite{lukasczyk2017viscous} at $t{=}51$ (9~regions) and $t{=}52$ (5~regions); counts exclude the unassigned manifold (\cref{sec:formulation}). (b)~Exploded view revealing all regions individually, labeled S1--S9 and T1--T5. (c)~Region coupling~$\xi$ for $t{=}51{\to}52$, with rows normalized for display: each bar shows how a source region distributes mass across target regions. S1 and S3 merge into~T1; S5, S6, and S8 consolidate into~T4; S4 persists as~T2; S2 and S9 split across multiple targets (dashed outlines). Entries below $0.005$ are not drawn, which leaves 1.6 targets per source region on average against 2.3 entries above $10^{-12}$ in the coupling itself, so the panel reads as sparser than $\xi$ is. (d)~Multidimensional scaling (MDS) of the 15 ensemble runs under WD, GWD, FGW, and MS-COOT, colored by the smoothing length $h$ of the particle simulation, its interaction radius (0.20, 0.30, 0.44; fine, medium, coarse in the contest's terms). Only MS-COOT separates the three groups.}
    \label{fig:teaser}
\end{figure*}

\section{Introduction}
\label{sec:introduction}
Scientific data arising from simulations, experiments, and observations in many domains are naturally represented as scalar fields. 
A central task in topology-based scientific visualization is to extract and compare meaningful structures within these fields, such as vortices in fluid flows~\cite{laney2006understanding}, burning cells in combustion~\cite{bremer2009analyzing}, or frontal boundaries in atmospheric dynamics~\cite{LanGamelin2024,LiChatterjee2025}. 
These structures are often characterized by their topological organization, motivating the use of descriptors that encode critical features and their relationships~\cite{heine2016survey,yan2021scalar}. 
Consequently, defining principled distances and correspondences between such descriptors is essential for tasks including comparison, clustering, and temporal analysis of scalar data.

Among such descriptors, the Morse-Smale (MS) complex~\cite{edelsbrunner2003hierarchical, gyulassy2008practical} provides a rich structural representation by decomposing the domain into regions of uniform gradient flow. 
It captures critical points as feature representatives and separatrices that define region boundaries, explicitly encoding both spatial embedding and adjacency. 
In contrast, persistence diagrams~\cite{edelsbrunner2002topological} summarize features as unordered birth-death pairs, and merge trees~\cite{carr2003computing} encode hierarchical connectivity of sublevel sets. 
While effective for measuring feature significance or hierarchy, these representations do not explicitly capture region-level structure and spatial relationships, which are central to many analysis tasks.

Comparing MS complexes, therefore, requires both a meaningful distance and a structural correspondence that respects this richer organization. 
Prior work has applied optimal transport (OT) to graph representations of MS complexes by considering their 1-skeletons~\cite{FengHuangJu2013}, matching critical points via a transport plan~\cite{li2023comparing}. While effective at the level of critical points (CPs), such approaches produce a single coupling indexed by critical points; a region correspondence derived from it by aggregating over boundary critical points has neither the region marginals nor a region objective behind it, and is not what the method optimizes.

To address this gap, we model the MS complex as a hypergraph, where critical points correspond to nodes and regions correspond to hyperedges incident to their boundary CPs. This formulation aligns naturally with co-optimal transport (COOT)~\cite{redko2020co}, which jointly optimizes separate couplings over different entity types. Building on recent extensions of COOT to hypergraphs~\cite{chowdhury2024hypergraph}, we introduce MS-COOT, a distance that simultaneously matches critical points and regions, enabling structure-aware comparison of MS complexes beyond graph-based representations. 
Our contributions are:
\begin{itemize}[noitemsep,leftmargin=*]

\item \textbf{MS-COOT: a hypergraph-based distance for MS complexes.} 
We formulate the Morse-Smale complex as a measure hypernetwork and introduce a co-optimal transport distance that jointly matches critical points and regions.

\item \textbf{Region-level correspondence and analysis.} 
MS-COOT produces an explicit region coupling $\xi$ in addition to the critical point coupling $\pi$. 
To the best of our knowledge, this is the first general distance-based formulation for MS complex comparison that produces region-level correspondences. 
This formulation enables direct identification of region-level events such as splitting and merging.

\item \textbf{Evaluation across datasets and tasks.} 
We evaluate MS-COOT on 2D and 3D scalar data, including time-varying simulations, surface meshes, and volumetric data. It reveals region-level structural changes that graph-based distances do not, achieves competitive accuracy in shape classification, and improves discrimination of simulation parameters.

\end{itemize}

\section{Related Work}
\label{sec:related}
We review approaches for comparing topological descriptors, including persistence diagrams, tree-based structures, and graph-based representations of Morse(-Smale) complexes. We then review topology-based tracking methods, which produce feature correspondences over time, and the optimal transport formulations used in these settings; see~\cite{heine2016survey,yan2021scalar} for surveys.

\para{Persistence diagrams.} A persistence diagram summarizes a filtration as a multiset of birth-death pairs~\cite{Dey2022,Oudot2015,edelsbrunner2010computational}. Distances between diagrams are defined through partial matchings of these pairs, with unmatched pairs sent to the diagonal: the Bottleneck distance minimizes the largest displacement of any matched pair, while $p$-Wasserstein distances minimize an aggregate displacement. The Bottleneck distance is stable under perturbations of the scalar function measured in the supremum norm~\cite{CohenSteiner2007}, and both distances are efficiently computable, which has made them a standard tool for comparing scalar fields. Vectorizations such as persistence images~\cite{adams2017persistence} further map diagrams into a fixed-dimensional space suitable for downstream learning. These methods characterize features through birth-death pairs, focusing on their persistence while abstracting away spatial embedding and connectivity, so a matching between diagrams identifies features of comparable significance, not corresponding structures in the domain.

\para{Tree-based descriptors.} Merge trees and contour trees~\cite{carr2003computing} encode the nesting of (sub-)level sets and have been extensively studied for comparison. Edit distances adapt tree-editing formulations to this setting, including constrained edit distances on branch decompositions~\cite{sridharamurthy2020edit}, branch-decomposition-independent variants~\cite{wetzels2022branch}, and deformation-based formulations~\cite{wetzels2022deformation}. 
Interleaving distances compare trees through shifted maps between them~\cite{morozov2013interleaving}, with intrinsic~\cite{Gasparovic2025} and geometry-aware~\cite{YanMasood2023} variants. Other formulations define Wasserstein distances, geodesics, and barycenters over merge trees~\cite{pont2022wasserstein}, train graph neural networks to approximate such comparisons at scale~\cite{QinFasy2025}, or define distances through direct structural matching~\cite{beketayev2014measuring}. 
These approaches capture hierarchical relationships induced by scalar filtrations.
More recently, Pont and Garth~\cite{pont2026region} incorporate spatial information from merge-tree branch regions into Wasserstein matching of persistence pairs.
These regions, which may be spatially disconnected, are associated with individual persistence pairs.
In contrast, MS-COOT represents Morse--Smale regions and critical points as distinct entities and jointly optimizes their correspondences through separate, structurally coupled transport plans.

\para{Graph-based MS complex comparison.} Previous work compares Morse(-Smale) complexes via their 1-skeletons~\cite{FengHuangJu2013,ThomasNatarajan2013}, representing critical points as nodes and separatrices as edges. Optimal transport distances have been applied in this setting for tasks such as feature correspondence and temporal analysis~\cite{li2023comparing}. The 1-skeleton records which critical points are connected and how the separatrices are embedded, and is therefore sufficient to support matching at the level of critical points. The grid samples bounded by those separatrices are not elements of this representation, so neither the distance nor the coupling computed on the 1-skeleton is indexed by them.

\para{Topology-based feature tracking.} A separate line of work seeks correspondences between features across time steps rather than a distance between descriptors. 
Saikia and Weinkauf~\cite{SaikiaWeinkauf2017} track features globally in time-dependent scalar fields using merge tree similarity. 
Soler et al.~\cite{soler2018lifted} combine Wasserstein matching of persistence diagrams with geometrical information for fast and robust topology tracking, and Li et al.~\cite{li2023flexible} formulate tracking probabilistically using partial optimal transport, allowing features to appear and disappear. 
Le Thanh and Weinkauf~\cite{LeThanhWeinkauf2025} compare several edit-distance-based tracking methods on time-varying data. In application settings, Lukasczyk et al.~\cite{lukasczyk2017viscous} build tracking graphs over feature volumes to analyze viscous fingering, and related pipelines have been developed for atmospheric structures~\cite{LiChatterjee2025}. These methods produce feature-level correspondences, and in some cases correspondences between spatially extended structures, which makes them the closest prior work to the region coupling we introduce. They differ from our setting in two respects. First, they are tracking pipelines, and the correspondence they produce is not necessarily accompanied by a distance usable for clustering, classification, or ensemble comparison. Second, where correspondences between spatial structures arise, they are obtained from spatial overlap or feature similarity along a time series, not from a matching problem posed jointly over the critical points and the region sets of an arbitrary pair of complexes.

\para{Optimal transport for topological structures.} Optimal transport (OT) has been widely applied to compare topological descriptors, including merge trees and graph-based representations of Morse-Smale complexes~\cite{LiPalandeYan2023,li2023flexible,li2023comparing}. These approaches typically rely on OT variants such as WD, GWD, and FGW (see \cref{sec:bg_ot}), which optimize a single transport coupling to align elements between two structures. Co-Optimal Transport (COOT)~\cite{redko2020co} extends this framework by jointly optimizing two couplings over different entity types, and has been further generalized to hypergraph-structured data~\cite{chowdhury2024hypergraph}. Neither has been instantiated for Morse-Smale complexes, which requires defining the node measure, the hyperedge measure, and the node-hyperedge relational function in a way that reflects the structure of the complex.

\para{Limitation and gap.}
Existing approaches compare scalar fields through topological descriptors, graph-based representations, or spatial feature tracking.
Although region information can be incorporated into these comparisons, existing distance formulations do not explicitly optimize correspondences between MS regions.
In particular, while the embedded 1-skeleton determines the MS region decomposition, critical-point-based distances and couplings do not directly characterize changes in the regions themselves.
A low-persistence critical-point cancellation, for example, may merge spatially extensive regions, illustrating the distinction between changes in critical-point structure and region decomposition.
Feature-tracking methods establish spatial correspondences but do not generally formulate a distance over arbitrary pairs of MS complexes.
To address this gap, we represent MS complexes as hypergraphs and introduce a co-optimal transport distance that jointly matches critical points and regions through separate, coupled transport plans.

\begin{figure*}[t]
    \centering
    \vspace{-2mm}
    \includegraphics[width=0.90\textwidth]{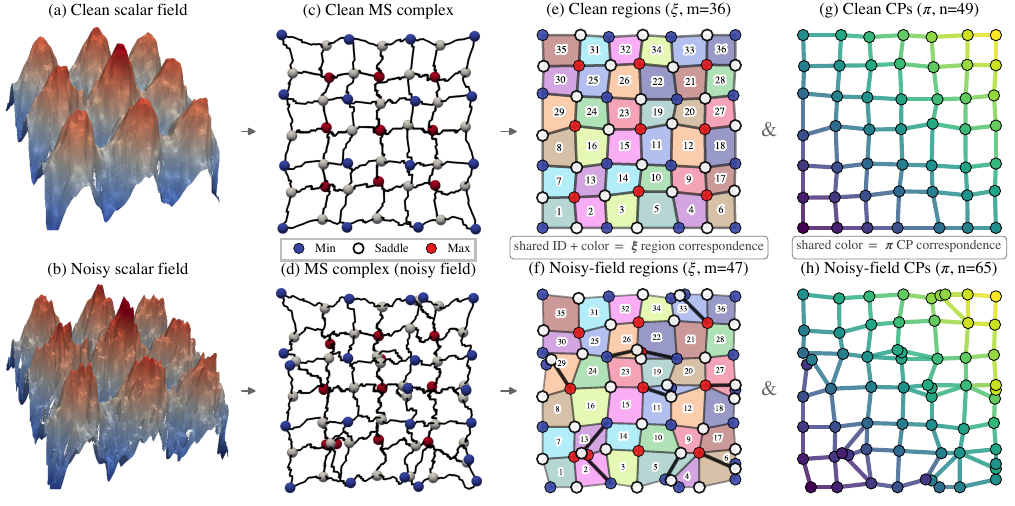}
    \vspace{-4mm}
    \caption{\textbf{MS-COOT simultaneously matches critical points and regions between two Morse-Smale complexes.} 
(a,\,b)~Clean and noisy sinusoidal scalar fields. 
(c,\,d)~Morse-Smale complexes extracted by TTK, showing critical points (CPs: minima, saddles, maxima) and separatrices. 
(e,\,f)~Region coupling~$\xi$: each region of the noisy field in~(f) inherits the color and ID of its best-matched region in the clean field~(e). When a single clean-field region corresponds to multiple noisy-field regions, these share the same color and ID, indicating noise-induced splits (bold separatrices highlight the new boundaries). 
(g,\,h) CP coupling $\pi$: each clean CP in (g) is colored by its position along the domain diagonal; each noisy-field CP in (h) inherits the color of its matched counterpart, so a correct matching preserves the spatial gradient~\cite{ovsjanikov2012functional,li2023comparing}. Points on the same anti-diagonal share a color. This illustration uses $\alpha = 0.001$;~\Cref{sec:appendix_measures,sec:appendix_cp_matching} evaluate it quantitatively.}
    \label{fig:sinusoidal}
    \vspace{-4mm}
\end{figure*}

\section{Background}
\label{sec:background}

We review Morse-Smale (MS) complexes (\cref{sec:bg_ms}), hypergraphs and measure hypernetworks (\cref{sec:bg_hypergraph}), and optimal transport (OT) formulations (\cref{sec:bg_ot}) that underpin our method. 

\subsection{Morse-Smale Complexes}
\label{sec:bg_ms}

Let \(M\) be a smooth manifold and \(f : M \to \mathbb{R}\) a Morse function. 
A point \(x \in M\) is \emph{critical} if \(\nabla f(x)=0\), and \emph{regular} otherwise. 
At a regular point, integrating the gradient field defines an integral line, i.e., a maximal curve whose tangent agrees with \(\nabla f\). 
Each integral line begins and ends at critical points. 
Critical points are classified as minima, saddles, or maxima according to the local behavior of \(f\). 

The Morse-Smale (MS) complex decomposes \(M\) according to this gradient flow behavior~\cite{edelsbrunner2003hierarchical,forman1998morse,milnor1963morse}. 
The resulting regions exhibit uniform flow, with all regular points in each region lying on integral lines that originate from the same minimum and terminate at the same maximum.
In 2D, these regions are bounded by separatrices, which are integral lines connecting saddles to other critical points. In 3D, region boundaries also include two-dimensional separating manifolds formed by integral lines associated with saddles.

\Cref{fig:sinusoidal}(a--d) illustrates this construction. 
Given a scalar field (a), the MS complex (c) captures its critical points and separatrices and partitions the domain into regions. 
Under perturbation (b), the resulting MS complex (d) exhibits structural changes in both critical points and region decomposition.

In practice, discretized approximations of MS complexes are commonly computed in piecewise-linear (PL) settings on triangulated domains, with established constructions in both 2D~\cite{edelsbrunner2003hierarchical} and 3D~\cite{edelsbrunner2003morse3d}. 
Critical points can be paired by topological persistence~\cite{edelsbrunner2002topological}: each pair records the scalar values $(b_i, d_i)$ at which a feature is created and destroyed, and its persistence $d_i - b_i$ measures the significance of that feature. 
Cancelling pairs below a persistence threshold removes small-scale structure and yields a simplified complex, which is the mechanism we use to control the granularity of the complexes being compared. 
To compare MS complexes, it is useful to explicitly represent relationships between critical points and regions. 
Each region is bounded by multiple critical points, and each critical point may be incident to multiple regions, forming a higher-order connectivity structure. 
This observation suggests a hypergraph representation of MS complexes, where critical points correspond to nodes and regions correspond to hyperedges incident to their boundary critical points.

\subsection{Hypergraphs and Measure Hypernetworks}
\label{sec:bg_hypergraph}

A hypergraph is a pair $(V, E)$, where $V$ is a set of nodes and $E \subseteq \mathcal{P}(V)$ is a collection of subsets of $V$, called hyperedges; here $\mathcal{P}(V)$ denotes the power set of $V$. 
Unlike graphs, hyperedges may connect more than two nodes, allowing representation of higher-order relationships.

To compare hypergraphs in a structured manner, Chowdhury et al.~\cite{chowdhury2024hypergraph} introduce the notion of a \emph{measure hypernetwork}, which enriches a hypergraph with additional structure. 
Formally, a measure hypernetwork is a tuple
\[
\mathcal{H} = (V, E, \mu, \nu, \omega),
\]
where:
\begin{itemize}[noitemsep,leftmargin=*]
    \item $V$ is a finite set of nodes,
    \item $E$ is a finite set of hyperedges,
    \item $\mu \in \Delta(V)$ is a probability measure on nodes,
    \item $\nu \in \Delta(E)$ is a probability measure on hyperedges, and
    \item $\omega : V \times E \to \mathbb{R}$ is a function encoding relationships between nodes and hyperedges.
\end{itemize}

Here, $\Delta(V)$ and $\Delta(E)$ denote the probability simplices over $V$ and $E$, respectively. 
The function $\omega(v,e)$ quantifies how node $v$ is related to hyperedge $e$, capturing relational structure within the hypergraph.

This formulation extends classical hypergraphs by incorporating both \emph{probability measures} and \emph{relational information}, providing a unified representation suitable for optimal transport-based comparison.

\subsection{Optimal Transport for Structured Comparison}
\label{sec:bg_ot}

Optimal transport (OT)~\cite{peyre2019computational,peyre2016gromov, vayer2019sliced, meng2025efficient,tran2023unbalanced} provides a framework for comparing structured data by transporting probability mass between two distributions. 
Let $\mathcal{X}$ and $\mathcal{Y}$ be finite sets with probability measures $a \in \Delta(\mathcal{X})$ and $b \in \Delta(\mathcal{Y})$. 
A transport coupling is a joint distribution $\pi \in \Pi(a,b)$, where
\begin{equation*}
\resizebox{0.98\columnwidth}{!}{$\displaystyle
\Pi(a,b) = \left\{
\pi \in \mathbb{R}_{+}^{|\mathcal{X}|\times|\mathcal{Y}|}
\;\middle|\;
\sum_{j} \pi[i,j] = a[i],\;
\sum_{i} \pi[i,j] = b[j]
\right\}.
$}
\end{equation*}
The coupling $\pi[i,j]$ specifies how much probability mass is transported from element $i \in \mathcal{X}$ (source) to element $j \in \mathcal{Y}$ (target), subject to conservation of mass. 
Given a cost matrix $L \in \mathbb{R}^{|\mathcal{X}|\times|\mathcal{Y}|}$ scoring the assignment of $i$ to $j$, the Wasserstein distance (WD) minimizes the total transport cost,
\[
\mathrm{WD}(a,b) = \min_{\pi \in \Pi(a,b)} \sum_{i,j} L[i,j] \, \pi[i,j],
\]
which is a linear program. 
Adding an entropic penalty $-\varepsilon H(\pi)$ makes the problem strictly convex and solvable by iterative matrix scaling with the Sinkhorn algorithm~\cite{cuturi2013sinkhorn}, at the price of a denser coupling.

\para{Gromov-Wasserstein and fused variants.}
When the two sets are not embedded in a common space, no cost matrix $L$ is available. 
The Gromov-Wasserstein distance (GWD)~\cite{memoli2011gromov} instead compares intra-domain distance matrices $D_\mathcal{X}$ and $D_\mathcal{Y}$,
\[
\min_{\pi \in \Pi(a,b)} \sum_{i,i',j,j'} \big|D_\mathcal{X}[i,i'] - D_\mathcal{Y}[j,j']\big|^2 \, \pi[i,j] \, \pi[i',j'],
\]
so that pairs of elements at similar internal distances are matched together. 
The fused Gromov-Wasserstein distance (FGW)~\cite{titouan2019optimal} interpolates between the two, combining the linear cost of WD with the quadratic term of GWD through a parameter $\beta \in [0,1]$. 
These three distances serve as our baselines in \cref{sec:results}. 
Note that $\pi$ appears twice in the GWD objective: a single coupling over a single type of entity carries the entire comparison.

\para{Co-Optimal Transport (COOT).}
Co-Optimal Transport (COOT)~\cite{redko2020co} extends OT to jointly compare two types of entities. 
Consider two datasets consisting of entities of type $A$ and $B$, with relationships encoded by matrices $\Omega_1 \in \mathbb{R}^{|A_1| \times |B_1|}$ and $\Omega_2 \in \mathbb{R}^{|A_2| \times |B_2|}$. 
Let $\mu_1 \in \Delta(A_1)$ and $\mu_2 \in \Delta(A_2)$ be distributions over entities of type $A$, and $\nu_1 \in \Delta(B_1)$ and $\nu_2 \in \Delta(B_2)$ over entities of type $B$.

COOT introduces two transport problems:
\[
\pi \in \Pi(\mu_1,\mu_2), 
\quad 
\xi \in \Pi(\nu_1,\nu_2),
\]
where $\pi$ transports mass between entities of type $A$ and $\xi$ transports mass between entities of type $B$. 
The objective
\[
\sum_{i,j,k,l} |\Omega_1[i,k] - \Omega_2[j,l]|^2 \, \pi[i,j] \, \xi[k,l]
\]
couples these two transport plans, ensuring that aligned entities preserve their relational structure.

In contrast to GWD, the two couplings range over different entity types, so the objective is bilinear rather than quadratic in a single variable. 
Fixing one coupling leaves a linear problem in the other, which suggests a block coordinate descent scheme alternating updates of $\pi$ and $\xi$, each update being an entropic OT problem solved with Sinkhorn~\cite{redko2020co}. 
We describe our instantiation of this scheme in \cref{sec:optimization}.

This formulation is well-suited to MS complexes, where critical points and regions form two distinct entity types. 
As illustrated in \cref{fig:sinusoidal}(e--h), the coupling $\pi$ transports mass between critical points, while $\xi$ transports mass between regions.

\begin{figure*}[t!]
\centering
\resizebox{0.95\textwidth}{!}{%
\begin{tikzpicture}[
    inputbox/.style={
        rectangle, draw=gray!65, very thick, 
        text width=3.2cm, minimum height=1.5cm, 
        align=center, rounded corners=4pt, font=\normalsize
    },
    compbox/.style={
        rectangle, draw=gray!65, very thick, 
        text width=5.0cm, minimum height=1.8cm, 
        align=center, rounded corners=4pt, font=\normalsize
    },
    solverbox/.style={
        rectangle, draw=gray!65, very thick, 
        text width=4.0cm, minimum height=1.8cm, 
        align=center, rounded corners=4pt, font=\normalsize
    },
    outbox/.style={
        rectangle, draw=gray!65, thick, 
        text width=4.2cm, minimum height=1.1cm, 
        align=center, rounded corners=3pt, font=\normalsize
    },
    mainarrow/.style={
        -{Stealth[length=5pt, width=4pt]}, 
        line width=1.2pt, arrowcolor
    },
    brancharrow/.style={
        -{Stealth[length=4pt, width=3.5pt]}, 
        line width=0.9pt, arrowcolor!80
    },
    trunk/.style={
        line width=0.9pt, arrowcolor!80
    },
]

    \node[inputbox, fill=inputgray] (input) at (0, 1.3) {
        \textbf{(a) Input}\\[0.1em]
        Scalar fields $f,\, g$
    };

    \node[inputbox, fill=ttkblue] (ttk) at (0, -1.3) {
        \textbf{(b) Morse-Smale}\\
        \textbf{Complex}\\[0.05em]
        CPs, \mbox{separatrices},\\
        regions
    };

    \draw[mainarrow] (input.south) -- (ttk.north);

    \node[compbox, fill=omegagreen] (omega) at (5.0, 1.3) {
        \textbf{(c) \mbox{Hypernetwork} Function $\boldsymbol{\omega}$}\\[0.12em]
        Graph with\\
        region centers (RCs)\\[0.02em]
        {\small $\omega[i,j] = d_G(\mathrm{CP}_i,\, \mathrm{RC}_j)$}
    };

    \node[compbox, fill=measurepurple] (measure) at (5.0, -1.3) {
        \textbf{(d) Measures $\boldsymbol{\mu}$ and $\boldsymbol{\nu}$}\\[0.12em]
        $\mu$: persistence-based probability\\[0.02em]
        $\nu$: aggregated from $\mu$\; (regions)
    };

    \coordinate (forkpt) at ($(ttk.east)+(0.35, 0)$);
    \coordinate (trunk_top) at (forkpt |- omega.west);
    \coordinate (trunk_bot) at (forkpt |- measure.west);

    \draw[trunk] (ttk.east) -- (forkpt);
    \draw[trunk] (trunk_top) -- (trunk_bot);
    \draw[brancharrow] (trunk_top) -- (omega.west);
    \draw[brancharrow] (trunk_bot) -- (measure.west);
    \fill[arrowcolor] (forkpt) circle (2pt);

    \node[solverbox, fill=solverorange] (solver) at (10.6, 0) {
        \textbf{(e) \mbox{MS-COOT} Solver}\\[0.15em]
        Co-Optimal Transport
    };

    \coordinate (merge_x) at ($(solver.west)+(-0.5, 0)$);
    \coordinate (merge_top) at (merge_x |- omega.east);
    \coordinate (merge_bot) at (merge_x |- measure.east);
    \coordinate (merge_mid) at (merge_x);

    \draw[trunk] (omega.east) -- (merge_top);
    \draw[trunk] (measure.east) -- (merge_bot);
    \draw[trunk] (merge_top) -- (merge_bot);
    \draw[brancharrow] (merge_mid) -- (solver.west);
    \fill[arrowcolor] (merge_mid) circle (2pt);

    \node[outbox, fill=piblue, minimum height=1.1cm] (pi) at (15.4, 1.5) {
        \textbf{(f) CP coupling $\boldsymbol{\pi}$}\\[0.02em]
        $\pi \in \Pi(\mu_f,\, \mu_g)$
    };

    \node[outbox, fill=xiorange, draw=xiaccent, line width=1.8pt, minimum height=1.1cm] (xi) at (15.4, 0) {
        \textbf{(g) Region coupling $\boldsymbol{\xi}$}\\[0.02em]
        $\xi \in \Pi(\nu_f,\, \nu_g)$
    };

    \node[outbox, fill=distred, minimum height=1.1cm] (dist) at (15.4, -1.5) {
        \textbf{(h) \mbox{MS-COOT} distance}\\[0.02em]
        $d_{\mathrm{MS\text{-}COOT}}(\mathcal{H}_f,\, \mathcal{H}_g)$
    };

    \coordinate (outfork) at ($(solver.east)+(0.3, 0)$);
    \coordinate (out_top) at (outfork |- pi.west);
    \coordinate (out_bot) at (outfork |- dist.west);

    \draw[trunk] (solver.east) -- (outfork);
    \draw[trunk] (out_top) -- (out_bot);
    \draw[brancharrow] (out_top) -- (pi.west);
    \draw[brancharrow] (outfork) -- (xi.west);
    \draw[brancharrow] (out_bot) -- (dist.west);
    \fill[arrowcolor] (outfork) circle (2pt);

\end{tikzpicture}%
}
\vspace{-2mm}
\caption{
\textbf{MS-COOT workflow.}
Given two scalar fields $f$ and~$g$ in~(a), we compute their
Morse-Smale complexes in~(b) and represent each as a measure
hypernetwork~$\mathcal{H}$. Two inputs are constructed for the
MS-COOT solver: the hypernetwork function $\omega$ in~(c),
encoding CP-to-region proximity via shortest-path distances
$d_G$ on an augmented graph with region
centers~(RCs), and persistence-based probability measures
$\mu$,~$\nu$ in~(d). The MS-COOT solver in~(e) jointly optimizes
CP coupling $\pi$ in~(f) and region coupling $\xi$ in~(g) via
co-optimal transport, yielding the distance
$d_{\mathrm{MS\text{-}COOT}}(\mathcal{H}_f, \mathcal{H}_g)$ in~(h).
Region coupling $\xi$ is unique to our method, enabling
region-level correspondence.
}
\label{fig:workflow}
\vspace{-3mm}
\end{figure*}

\section{Method}
\label{sec:method}

We introduce MS-COOT, a co-optimal transport formulation for comparing Morse-Smale complexes. Our key idea is to represent each MS complex as a hypergraph-structured object and jointly match its critical points and regions using two transport couplings: a critical point coupling $\pi$ and a region coupling $\xi$. This enables unified comparison of feature-level and region-level structures.

\cref{fig:workflow} provides an overview of the MS-COOT pipeline. We instantiate this formulation using three components: a hypernetwork function $\omega$ encoding relationships between critical points and regions, probability measures $\mu$ and $\nu$ over nodes and hyperedges, and a sample cost term penalizing inconsistency of critical point attributes such as types. The resulting objective is optimized via block coordinate descent with entropic regularization. The remainder of this section details the formulation (\cref{sec:formulation}), hypernetwork construction (\cref{sec:hypernetwork}), measures (\cref{sec:measures}), examples (\cref{sec:controlled}), and optimization (\cref{sec:optimization}). A summary of notation is provided in~\Cref{sec:appendix_notation}.

\subsection{MS-COOT Distance}
\label{sec:formulation}
Let $f$ and $g$ be two scalar fields on $M$. We encode each Morse-Smale (MS) complex as a measure hypernetwork by mapping its structural elements to the hypernetwork representation introduced in \cref{sec:bg_hypergraph}. Specifically, we associate critical points with nodes and regions (cells) with hyperedges, yielding
\[
\mathcal{H}_f = (V_f, E_f, \mu_f, \nu_f, \omega_f), 
\qquad
\mathcal{H}_g = (V_g, E_g, \mu_g, \nu_g, \omega_g),
\]
where $V$ is the set of critical points, $E$ is the set of regions, $\mu$ and $\nu$ are probability measures on nodes and hyperedges, and $\omega : V \times E \to \mathbb{R}$ encodes relationships between critical points and regions. Pixels or voxels left unassigned by the segmentation are grouped into a single catch-all region, which we exclude from $E$ because it does not correspond to an explicitly identified Morse–Smale region.

We define the MS-COOT distance as a co-optimal transport problem that jointly aligns nodes and hyperedges of these hypernetworks. This formulation builds on co-optimal transport~\cite{redko2020co} and its extension to hypergraph-structured data~\cite{chowdhury2024hypergraph}, with additional domain-specific components tailored to MS complexes.

\para{Definition.}
The MS-COOT distance between $\mathcal{H}_f$ and $\mathcal{H}_g$ is defined as
\begin{equation}
\label{eq:mscoot}
\begin{aligned}
d_{\mathrm{MS\text{-}COOT}}(\mathcal{H}_f, \mathcal{H}_g)
= \\
\min_{\pi,\, \xi} \;
\sum_{i,j,k,l}
  \bigl| \omega_f[i,k] &- \omega_g[j,l] \bigr|^2
  \pi[i,j] \, \xi[k,l] 
+\; \alpha \langle C, \pi \rangle.
\end{aligned}
\end{equation}
Here, $\pi \in \Pi(\mu_f, \mu_g)$ and $\xi \in \Pi(\nu_f, \nu_g)$ are transport couplings between critical points and regions, respectively, and $\Pi(\cdot,\cdot)$ denotes the set of admissible couplings defined in \cref{sec:bg_ot}. Both hypernetwork functions are rescaled to $[0,1]$ before the objective is formed, as described in \cref{sec:hypernetwork}.

\para{Structure term.}
The quadratic term in \cref{eq:mscoot} measures structural consistency between the two hypernetworks. It compares relational values $\omega_f[i,k]$ and $\omega_g[j,l]$ under the joint coupling $(\pi,\xi)$, encouraging that matched critical points and regions exhibit similar relationships. This corresponds to the standard co-optimal transport objective applied to $\omega_f$ and $\omega_g$.

\para{Critical-point sample cost.} The linear term $\alpha\langle C,\pi\rangle$ complements the relational term with information attached directly to critical points. The relational function $\omega$ describes how a critical point is situated relative to regions, but does not by itself encode attributes such as critical-point type. Our default sample cost therefore penalizes cross-type correspondences: \begin{equation} \label{eq:type_penalty} C_{\mathrm{type}}[i,j] = \mathbf{1}\!\left[ \mathrm{type}(\mathrm{CP}_i) \neq \mathrm{type}(\mathrm{CP}_j) \right]. \end{equation} More generally, $C$ can incorporate additional pointwise attributes when they are relevant to a particular comparison, such as normalized spatial distance or scalar-value difference between critical points. These terms provide information that is complementary to the relational structure encoded by $\omega$. We specify the exact form and weighting of $C$ for each experiment where it differs from the default.

\para{Outputs.}
The optimization yields a node coupling $\pi$, a hyperedge coupling $\xi$, and the distance $d_{\mathrm{MS\text{-}COOT}}(\mathcal{H}_f, \mathcal{H}_g)$. While existing OT-based methods produce node-level correspondences, the hyperedge coupling $\xi$ enables identification of region-level events such as splitting and merging.

The construction of $\omega$, $\mu$, and $\nu$ for MS complexes is described in the following subsections.


\begin{figure}[tb]
\centering
\includegraphics[width=0.8\columnwidth]{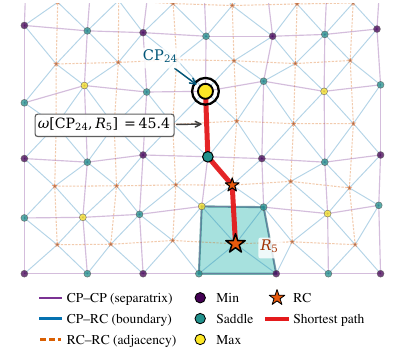}
\vspace{-2mm}
\caption{\textbf{Shortest-path example on the augmented graph $G$}, constructed from the clean sinusoidal dataset (\cref{fig:sinusoidal}(a)). Three edge types connect critical points and region centers (RCs): CP--CP edges along separatrices, CP--RC boundary edges, and RC--RC adjacency edges (see legend). The red path realizes $\omega[\mathrm{CP}_{24}, R_5] = 45.4$, the shortest-path distance from a critical point to the center of a region it does not bound. It traverses all three edge families in turn, illustrating how $\omega$ encodes graded proximity across the entire complex. Lengths are shown before the per-complex normalization of \cref{eq:normal}.}
\vspace{-3mm}
\label{fig:omega}
\end{figure}

\subsection{Hypernetwork Function}
\label{sec:hypernetwork}

In the hypernetwork formulation (\cref{sec:bg_hypergraph}), the relational
function $\omega : V \times E \to \mathbb{R}$ encodes relationships between
nodes and hyperedges. For an MS complex with critical points
$V=\{\mathrm{CP}_1,\dots,\mathrm{CP}_n\}$ and regions
$E=\{R_1,\dots,R_m\}$, we represent $\omega$ as an $n\times m$ matrix; we use
$n_f,m_f$ and $n_g,m_g$ when the two complexes must be distinguished.
Different hypernetwork functions can encode different aspects of these relationships, including binary incidence and shortest-path relations~\cite{chowdhury2024hypergraph}. 
Here, we use a distance-valued relation designed to capture both MS connectivity and geometric separation
between critical points and regions.

\para{Region centers.}
To define this relation, each region $R_j$ is represented by an auxiliary point,
its \emph{region center} $\mathrm{RC}_j$. Geometrically, a route from an external critical point to a region can
be viewed as first reaching its boundary through a boundary critical point or
a neighboring region, followed by travel within the region. This motivates an
intrinsic representative based on within-region geodesic distance. A natural
reference is the \emph{geodesic medoid}: among the pixel centers in a region,
the point minimizing the sum of geodesic distances to all other pixel centers.
This is the sample-restricted counterpart of the Fr\'echet median on a
geodesic metric space~\cite{arnaudon2013medians}. Fr\'echet medians on
Riemannian manifolds have established existence, consistency, and robustness
properties under suitable geometric conditions~\cite{arnaudon2013medians}. 
However, obtaining within-region distances is computationally heavy: A
single-source solve using the Fast Marching Method (FMM)~\cite{sethian1999fast,kimmel1998geodesic} over $N$ pixels costs $O(N\log N)$, so exhaustively testing
all $N$ candidate medoids costs $O(N^2\log N)$ with this construction.

To avoid the high cost of computing the geodesic medoid, we use the geometric centroid of the spatial samples within each region as an alternative to represent each region by default:
\begin{equation}
\label{eq:rc}
\mathrm{RC}_j = \frac{1}{|X_j|}\sum_{x\in X_j}x,
\end{equation}
where $X_j$ contains the ambient coordinates of the spatial samples assigned to $R_j$.
Depending on the data representation, these samples may correspond to pixel centers, voxel centers, mesh vertices, or other discrete spatial locations.
The centroid minimizes the sum of squared Euclidean distances to these samples and is computed in one pass in $O(N)$ work.
It has a simple stability property: for normalized spatial sample distributions $\eta$ and $\tilde{\eta}$ with centroids $c$ and $\tilde{c}$,
\begin{equation}
\label{eq:centroid_stability}
\|c-\tilde{c}\|_2 \leq W_1(\eta,\tilde{\eta}),
\end{equation}
where $W_1$ uses Euclidean ground distance~\cite{peyre2019computational}.
Thus, small transport perturbations of the sampled region geometry induce small centroid displacements.
However, this does not imply stability of the MS segmentation itself, nor does it prevent substantial changes under region splits or merges.

\para{Augmented graph.}
We construct a weighted graph $G=(V_G,E_G)$ (\cref{fig:omega}) with node set
\begin{equation}
\label{eq:augmented_graph}
V_G = V \cup \{\mathrm{RC}_1,\dots,\mathrm{RC}_m\},
\end{equation}
and three types of edges:
\begin{itemize}[noitemsep]
    \item CP--CP edges between critical points joined by a separatrix;
    \item CP--RC edges between each region center and its boundary critical points;
    \item RC--RC edges between the centers of adjacent regions.
\end{itemize}
All edges are weighted by the Euclidean distance between their endpoints in
ambient coordinates, independently of function values. The resulting graph
therefore combines the connectivity of the MS complex with geometric distances
between its critical points and region representatives.

\para{Definition of $\omega$.}
We define the hypernetwork function using shortest-path distances on $G$:
\begin{equation}
\label{eq:omega}
\omega[i,j] = d_G(\mathrm{CP}_i,\mathrm{RC}_j),
\quad i\in[n],\; j\in[m],
\end{equation}
where $d_G$ denotes the weighted shortest-path distance. In
\cref{fig:omega}, the path from $\mathrm{CP}_{24}$ to $\mathrm{RC}_5$
illustrates how CP--CP, CP--RC, and RC--RC edges jointly relate a critical
point to a non-incident region. Alternative definitions of $\omega$, including
incidence-, hop-, and Euclidean-based relations, are evaluated in
\cref{sec:ablation}.

We normalize each complex independently using its largest entry:
\begin{equation}
\label{eq:normal}
Z_f = \max_{i,k}\omega_f[i,k],
\qquad
Z_g = \max_{j,l}\omega_g[j,l],
\end{equation}
and rescale $\omega_f\leftarrow\omega_f/Z_f$ and
$\omega_g\leftarrow\omega_g/Z_g$, placing both matrices in $[0,1]$.
Each complex therefore retains the same representation across comparisons.

\para{Representative comparison.}
We evaluate the centroid against an exhaustive
FMM-based geodesic-medoid reference on 157 timesteps of the Vortex Street dataset (\cref{sec:vortex}).
With per-complex normalization, their relational matrices differ by
\[
\frac{\|\omega-\omega_{\mathrm{FMM}}\|_F}
     {\|\omega_{\mathrm{FMM}}\|_F},
\]
with a mean of $0.31\%$ and a range of $0.19\%$--$0.97\%$ over timesteps.
Computing all centroids takes $1.4$\,ms per timestep, compared with
$532$\,ms for the exhaustive geodesic reference, a $380$-fold reduction in
representative-computation time. Only 13 of 13{,}747 centroids lie outside their regions, ten of which belong to regions containing at most 11 pixels. 
These results indicate that the centroid can work as an efficient practical surrogate for the intrinsic representative on this dataset.

\subsection{Persistence-Based Measures}
\label{sec:measures}

The probability measures $\mu$ and $\nu$ determine the relative
importance of critical points and regions in the transport. We define
these measures to account for the significance of the corresponding
features, first constructing the node measure $\mu$ and then deriving
the region measure $\nu$ from its boundary critical points.

\para{Node measure $\mu$.}
Critical points may differ substantially in their significance, for
which persistence provides a natural measure. Rather than assigning
weights directly from persistence values, we use a persistence-image
(PI) representation~\cite{adams2017persistence}, following the
PI-based probability measure used for Reeb graph comparison
in~\cite{chambers2025reeb}.
Let $\mathcal{D}$ denote the persistence diagram, written in birth-persistence coordinates $(b,p)$ with $p=d-b$. For a paired critical point $\mathrm{CP}_i$, let $(b_i,p_i)$ denote the coordinates of its persistence pair. The PI construction forms a persistence surface using persistence-weighted Gaussians of bandwidth $\sigma$:
\begin{equation}
\label{eq:surface}
\rho_{\mathcal{D},\sigma}(x) \;=\; \sum_{(b_k, p_k) \in \mathcal{D}} p_k \, \mathcal{N}\bigl(x;\, (b_k, p_k),\, \sigma^2 I\bigr).
\end{equation}
We obtain the weight of $\mathrm{CP}_i$ by taking the inner product of this surface with a Gaussian centered at $(b_i,p_i)$ and normalizing:
\begin{equation}
\label{eq:mu}
\mu_i = \frac{w_i}{\sum_{k=1}^{n} w_k},
\qquad
w_i = \bigl\langle \rho_{\mathcal{D},\sigma},\; \mathcal{N}\bigl(\cdot\,;\, (b_i, p_i),\, \sigma^2 I\bigr) \bigr\rangle .
\end{equation}
Critical points without a finite persistence pair are assigned weight $1/n$, followed by renormalization of $\mu$. Numerical evaluation is described in~\Cref{sec:appendix_implementation}.

Because the persistence surface aggregates contributions from
all pairs in $\mathcal{D}$, each CP's weight depends on both its own
persistence and the persistence-weighted concentration of
nearby pairs in birth--persistence space.
The bandwidth $\sigma$ controls the extent of this interaction.
Consequently, a CP within a cluster of moderate-persistence
pairs may receive more weight than an isolated CP with
greater persistence, a tradeoff that need not be desirable
for every application.

\para{Hyperedge measure $\nu$.}
Regions do not have intrinsic persistence values, so we derive their weights from their boundary critical points:
\begin{equation}
\label{eq:nu}
\nu_j = \frac{\sum_{\mathrm{CP}_i \in \partial R_j} \mu_i}
  {\sum_{l=1}^{m} \sum_{\mathrm{CP}_i \in \partial R_l} \mu_i},
\end{equation}
where $\partial R_j$ denotes the set of boundary critical points of region $R_j$. The choice of $\mu$ therefore influences both critical-point and region transport through $\mu$ and $\nu$.

\para{Comparison with alternative measures.}
Previous works commonly use uniform node weights
($\mu_i=1/n$)~\cite{li2023comparing,li2023flexible,titouan2019optimal},
treating all critical points equally and leaving noise
suppression to preprocessing.
Raw-persistence weighting instead distinguishes feature
significance, but changes in persistence pairing can create
weight imbalances between spatially corresponding CPs,
leading to mismatched transport.
PI-based weighting provides a smoothed alternative that
retains feature significance while mitigating this sensitivity.
The controlled examples in \cref{sec:controlled} demonstrate
both behaviors: PI weighting distinguishes low- and
high-prominence perturbations while reducing mismatched
transport compared with raw-persistence weighting.

Other smoothing strategies could serve similar purposes.
We adopt PI as an established construction with diagram-level
stability results~\cite{adams2017persistence} and precedent
in stable Reeb-graph OT~\cite{chambers2025reeb},
without assuming these guarantees extend to MS-COOT.
We therefore treat PI as a practical default, with probability
measures and their parameters remaining task- and
data-dependent choices.

\begin{figure}[tb]
\centering
\includegraphics[width=0.96\columnwidth]{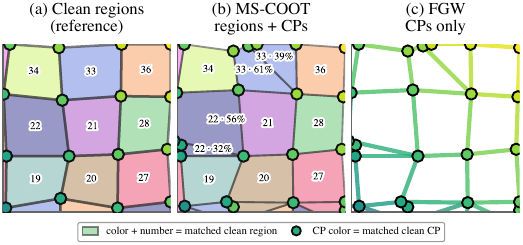}
\vspace{-3mm}
\caption{\textbf{Region and critical-point correspondence under noise.}
A window from the clean/noisy pair in \cref{fig:sinusoidal}, containing nine reference regions and sixteen reference critical points.
(a)~Clean reference.
(b)~MS-COOT with persistence-image node weights. Each noisy region is colored and labeled by its matched reference region under $\xi$, and each critical point is colored by its match under $\pi$. When noise splits a reference region, the labels report the mass transported to the corresponding pieces.
(c)~FGW with the same node weights. Critical points are colored by their matches under the FGW coupling; regions are uncolored because FGW does not compute a region coupling.}
\label{fig:controlled}
\vspace{-2mm}
\end{figure}

  \begin{figure}[tb]                    
  \centering
  \includegraphics[width=0.96\columnwidth]{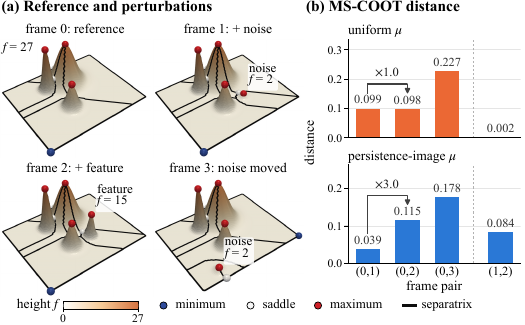}
  \vspace{-2mm}                         
\caption{\textbf{Noise and feature significance under different node measures.}
(a)~Four scalar fields on an $80 \times 100$ grid, with scalar value encoded by surface height and color, after persistence simplification at 1\% of the scalar range. Frame 0 is the reference. Frame 1 adds a low peak ($f=2$), frame 2 raises the same peak to $f=15$, and frame 3 moves the low peak to a different location. Frames 1 and 2 have the same critical points and nearly identical separatrices.
(b)~MS-COOT distances for selected frame pairs using uniform (top) and persistence-image (bottom) node weights. The two plots use independent vertical scales. Uniform weights give nearly identical distances for the low and prominent added peaks, whereas persistence-image weights give $d(0,2)/d(0,1)=3.0$. Moving the low peak in frame 3 produces the largest distance from the reference under both measures.}
  \vspace{-3mm}
  \label{fig:emerge}
  \end{figure}
  
\begin{figure}[tb]
\centering
\includegraphics[width=0.96\columnwidth]{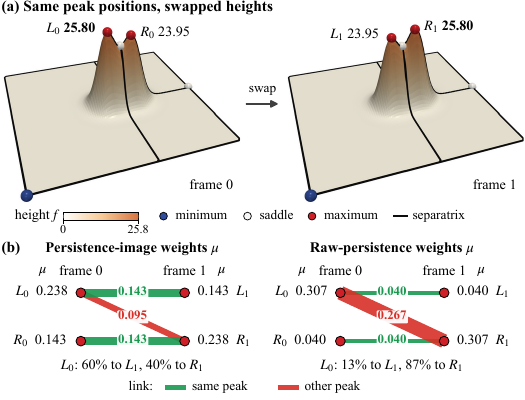}
\vspace{-4mm}
\caption{\textbf{Effect of persistence weighting on spatial correspondence.}
(a)~Two scalar fields formed by the same pair of peaks on a $100 \times 100$ grid. After persistence simplification at 1\% of the scalar range, the critical-point locations and separatrices are unchanged, while the two peak heights are exchanged, moving the global maximum from the left peak in frame 0 to the right peak in frame 1.
(b)~Critical-point coupling $\pi$ between the two maxima using persistence-image (left) and raw-persistence (right) node weights. Values beside the maxima give their normalized node weights $\mu$; link width and labels give transported mass. Green links connect spatially corresponding peaks, while red links indicate cross-peak transport. The minimum and saddles are omitted because they are matched to their corresponding critical points under both measures.}
\vspace{-3mm}
\label{fig:twopeak}
\end{figure}

\subsection{Controlled Examples}
\label{sec:controlled}

We use three controlled examples to examine the correspondence recovered by MS-COOT and its sensitivity to changes in feature significance and MS complex geometry.

\para{Region correspondence under noise.}
In the clean/noisy pair of \cref{fig:sinusoidal}, noise introduces additional critical points and subdivides existing regions. \Cref{fig:controlled} shows a window containing nine reference regions and sixteen reference critical points. With PI-based weights, $\xi$ identifies the noisy counterparts of all nine regions, including the splits of regions 22 and 33. For example, region 22 sends 56\% and 32\% of its mass to its two corresponding pieces. Over the complete domain, $\xi$ recovers 30 of 31 unsplit regions and four of five splits; see~\Cref{sec:appendix_measures} for more details.

At the critical-point level, MS-COOT matches 14 of the 16 points in this window, compared with all sixteen for WD and FGW under the same weights.
This difference partly reflects the evaluation configurations: WD and FGW directly penalize Euclidean displacement between CPs, whereas MS-COOT encodes spatial geometry through CP--region relationships in $\omega$, with a weak CP-type penalty ($\alpha=0.001$).
Thus, MS-COOT recovers most CP correspondences without a direct spatial matching cost, while also recovering region correspondences through $\xi$.
Spatial proximity can additionally be incorporated into $C$ when relevant to the task.

\para{Feature significance and geometric change.}
\Cref{fig:emerge} compares a reference field with three maxima against three perturbations. Frame 1 adds a low peak with $f=2$, representing a likely noise feature, and frame 2 raises the same peak to $f=15$. These two frames have the same critical-point configuration and nearly identical separatrices, but differ in the prominence of the added peak.

With uniform node weights, the distances from the reference are nearly equal,
$d(0,1)=0.099$ and $d(0,2)=0.098$, and the perturbed fields are nearly
indistinguishable ($d(1,2)=0.002$).
In contrast, PI-based weights yield $d(0,1)=0.039$ and $d(0,2)=0.115$,
a threefold difference, with $d(1,2)=0.084$.
Frame 3 relocates the low peak, altering the separatrices, region configuration,
and global minimum. Both measures assign it the largest distance from the reference.
Thus, MS-COOT responds to geometric reorganization under both measures,
while PI weighting additionally captures feature significance.

\para{Instability of persistence-based feature weights.}
Persistence assigned to spatially corresponding critical points can change substantially under small function-value perturbations when the persistence pairing switches. A related phenomenon is the vertical instability of persistence-based branch decompositions in merge-tree comparison: perturbations of nearly equal extrema can alter the branch hierarchy and disrupt feature correspondence~\cite{saikia2014extended,wetzels2022branch,bollen2023computing,wetzels2023taming}.

\Cref{fig:twopeak} illustrates the consequence for MS-COOT. The two peaks retain their positions and separatrices; but with small perturbations of function values, their heights, $25.80$ and $23.95$, are exchanged. The intended spatial correspondence remains left-to-left and right-to-right, while the maximum belonging to the high-persistence pair switches peaks. Under raw-persistence weighting, the weight of the left maximum changes from $0.307$ to $0.040$. The target left maximum can receive at most $0.040$, and the computed coupling sends the remaining $0.267$, or 87\% of the source mass, to the right maximum. Thus, the change in persistence pairing creates a marginal imbalance that forces transport away from the spatial correspondence.

With PI-based weights, the left-maximum weight becomes $0.143$, and the coupling sends $0.095$, or 40\% of its source mass, to the right maximum. 
PI-based weighting therefore mitigates the effect of pairing changes on the correspondence in this example. Uniform weights avoid this change in marginals, but do not distinguish the low and prominent peaks in \cref{fig:emerge}. These complementary observations motivate PI-based weighting as our default.

\subsection{Optimization Algorithm}
\label{sec:optimization}

Solving the COOT objective exactly is NP-hard, as it subsumes the quadratic assignment problem in the discrete unregularized setting~\cite{redko2020co,peyre2019computational}. Entropic regularization makes each subproblem of the alternating scheme below a matrix scaling problem solvable in polynomial time via the Sinkhorn algorithm~\cite{cuturi2013sinkhorn}; it does not make the joint problem tractable.

The MS-COOT objective (\cref{eq:mscoot}) is bilinear in the couplings $(\pi, \xi)$. Fixing one coupling reduces the optimization over the other to a convex problem, so we adopt a block coordinate descent (BCD) scheme~\cite{Tseng2001}, alternating between updates of $\pi$ and $\xi$. Each subproblem is an entropic optimal transport problem solved via Sinkhorn with regularization parameter $\varepsilon$, with the $\pi$-subproblem incorporating the additional linear cost $\alpha C$. If Sinkhorn fails to return a finite coupling, the subproblem is re-solved exactly at $\varepsilon = 0$ and the alternation continues; \cref{sec:implementation} reports how often this occurred. Each Sinkhorn iteration costs $O(n_f n_g)$ or $O(m_f m_g)$ depending on the coupling being updated, so a BCD iteration costs $O(K(n_f n_g + m_f m_g))$ in Sinkhorn steps plus $O(n_f n_g (m_f + m_g) + m_f m_g (n_f + n_g))$ to form the two cost matrices, where $K$ is the number of Sinkhorn iterations.

\para{Consistency of the two couplings.} The couplings are constrained only by their marginals and linked only through the structure term, so a region pair can carry mass in $\xi$ while the boundary critical points of the two regions carry little in $\pi$.~\Cref{sec:appendix_consistency} measures how often this occurs: on the time-varying datasets, between 8\% and 28\% of regions send less than half their boundary mass into the boundary of their matched region, and almost all of them are regions whose match in $\xi$ is itself uncertain.

\para{Convergence.} The alternation stops when the absolute change in the regularized objective between successive iterations falls below $10^{-6}$, or when the $\ell_1$ change in $\pi$ falls below $10^{-7}$, whichever occurs first, with a cap of $200$ block coordinate descent iterations and $200$ Sinkhorn iterations per subproblem. The objective monitored is the entropically regularized one, including its entropy terms, not the reported distance. Convergence of exact alternating minimization to a stationary point of the standard COOT objective is established by Redko et al.~\cite{redko2020co}. 
The additional linear term $\alpha \langle C, \pi \rangle$ preserves this property because it is linear in $\pi$ and constant in $\xi$, so the $\xi$-subproblem is unchanged and the $\pi$-subproblem remains a convex entropic OT problem with its cost shifted by $\alpha C$.
However, the exact alternating-minimization result does not directly guarantee stationary-point convergence for our finite-tolerance implementation, particularly when an unregularized fallback is used.



\para{Initialization.} Because the joint problem is non-convex, we run the alternation from several starting couplings, one deterministic and the rest random with fixed seeds, and retain the solution with the lowest objective;~\Cref{sec:appendix_implementation} specifies them.

\section{Experiments \& Results}
\label{sec:results}
We evaluate MS-COOT on four datasets spanning 2D time-varying scalar fields, 3D surface meshes, and 3D volumetric simulations. The 2D datasets are Vortex Street~\cite{popinet2003gerris} and Ionization Front~\cite{whalen2008ionization}.
A third 2D dataset, Heated Cylinder, is analyzed in~\Cref{sec:appendix_timevarying}.

\subsection{Setup}
\label{sec:implementation}
We extract MS complexes using ParaView 5.13.3~\cite{Ahrens2005ParaViewAE} with the TTK plugin~\cite{tierny2017topology}. For each instance, we compute the persistence diagram, simplify below a dataset-specific threshold, and extract the MS complex (critical points, separatrices, and regions). 
Thresholds are chosen by inspecting the elbow of each persistence curve, the point at which the number of surviving features stops decreasing rapidly~\cite{lukasczyk2020localized}, and are stated as a fraction of the scalar range: 
1\% for Vortex Street and TOSCA, 6\% for Viscous Finger, and 7\% for Ionization Front; see~\cref{sec:appendix_implementation} for details. 
From the simplified complex, we construct the augmented graph~$G$, compute the hypernetwork function~$\omega$ via shortest-path distances, and derive the measures~$\mu$ and~$\nu$. All transport computations use POT 0.9.1~\cite{flamary2021pot}. The exact fallback of \cref{sec:optimization} was needed on 5 of 435 Ionization Front pairs and on none of the 12{,}246 Vortex Street pairs. Parameter settings are given in~\Cref{sec:appendix_implementation}; dataset statistics and per-pair runtimes are in \cref{tab:implementation}. Our implementation is publicly available.\footnote{\url{https://github.com/gm3g11/MS-COOT}}

\para{Baselines.}
We compare MS-COOT with three OT-based baselines used for Morse(-Smale) complex comparison~\cite{li2023comparing}: Wasserstein distance (WD), Gromov--Wasserstein distance (GWD), and Fused Gromov--Wasserstein distance (FGW). Following Li et al.~\cite{li2023comparing}, each MS complex is represented by its 1-skeleton, with critical points as nodes and separatrices as edges weighted by the Euclidean distance between their endpoints, consistent with the CP--CP weights of \cref{sec:hypernetwork}. WD uses a cross-complex cost matrix $L$ of Euclidean distances between critical points on the time-varying datasets and the scalar cost of \cref{eq:scalar_cost} on TOSCA; GWD uses intra-complex shortest-path distances on the weighted 1-skeletons; and FGW combines both. All three baselines optimize a single critical-point coupling and do not model region correspondences.

We follow the unregularized optimization framework of Li et al.~\cite{li2023comparing} for the baselines: WD is solved by exact OT, while GWD and FGW use conditional gradient with an exact OT subproblem at each iteration. To separate the effect of critical-point weighting from that of the distance measure, on the time-varying datasets we evaluate each baseline with both uniform marginals and the persistence-based marginals. The main text reports the uniform baselines alongside both MS-COOT weighting settings; persistence-weighted baseline results are in~\Cref{sec:appendix_timevarying}.

\begin{figure*}[!ht]
  \centering
  \includegraphics[width=0.9\textwidth]{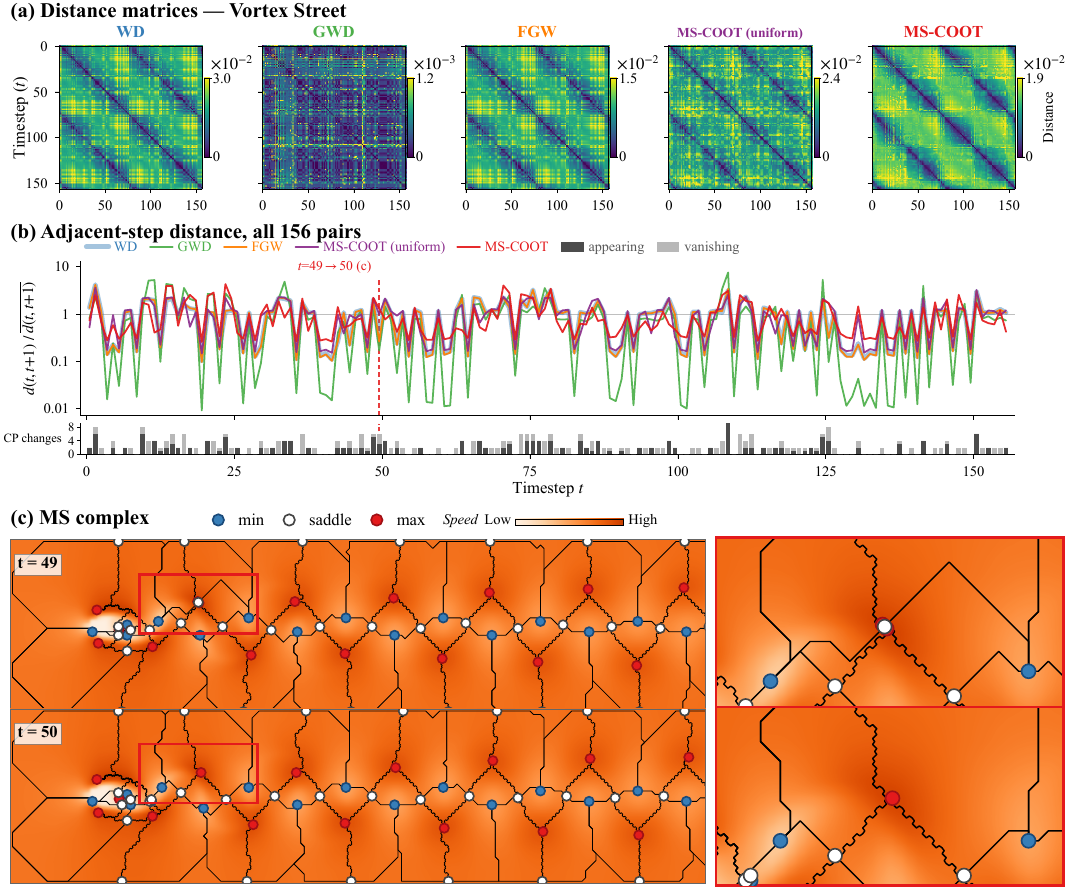}
  \vspace{-2mm}
\caption{\textbf{Vortex Street~\cite{popinet2003gerris}: shared periodicity and distinct sensitivity to region rearrangements.} Results for 157 consecutive timesteps. (a)~Pairwise distance matrices for WD, GWD, and FGW with uniform weights, and MS-COOT with uniform and persistence-image (PI) weights. Off-diagonal bands of low distances indicate a shedding period of 74--75 frames. (b)~Adjacent-step distances $d(t,t{+}1)$ normalized by each method's mean over the sequence (logarithmic scale). All five curves show similar overall trends, with WD and FGW nearly overlapping ($r=0.9996$). The histogram beneath shows critical-point births (dark) and deaths (light) between consecutive frames. At the marked step $49\to50$, WD, GWD, and FGW yield normalized distances of 0.27, 0.17, and 0.26, compared with 0.98 and 1.30 for uniform- and PI-weighted MS-COOT. (c)~MS complexes at $t=49$ and $50$, with the boxed area enlarged on the right. A saddle--maximum cancellation merges two neighboring regions, while a minimum--saddle pair appears near the lower-left corner of the enlarged view without a comparable region rearrangement. The critical-point count remains 117, while the region count decreases from 89 to 88.}
\vspace{-3mm}
\label{fig:vortexstreet}
\end{figure*}

\subsection{Vortex Street}
\label{sec:vortex}
The Vortex Street dataset contains 1501 timesteps of a 2D von K\'{a}rm\'{a}n vortex street simulated with the Gerris flow solver~\cite{popinet2003gerris,GuntherGrossTheisel2017,cgl}. We select 157 consecutive timesteps and compute all pairwise distances for each method (\cref{fig:vortexstreet}(a)). The flow exhibits periodic vortex shedding, as studied in prior work~\cite{LiPalandeYan2023,sridharamurthy2020edit,YanMasood2023}. Consecutive frames typically exhibit gradual downstream vortex motion, occasionally interrupted by substantial changes in the critical-point and region configurations.

\para{Periodicity.}
The distance matrices exhibit off-diagonal bands of low distances, indicating the recurrence of similar MS complexes after one shedding period. All methods show their strongest periodic similarity at a lag of 74 or 75 frames, with PI-weighted MS-COOT reaching its minimum at 75 frames. The banding is less pronounced for GWD. These results agree with the period estimated from the vortex-street geometry, $\tau=\lambda/U_c\approx75$ frames, using a same-sign vortex wavelength of $\lambda=28$ pixels and a convection speed of $U_c=0.375$ pixels per frame. The estimate is also consistent with the 75-frame period reported by Wetzels et al.~\cite{wetzels2022branch}, following Sridharamurthy et al.~\cite{sridharamurthy2020edit}.

\para{Temporal variations.}
\Cref{fig:vortexstreet}(b) compares adjacent-step distances normalized by each method's mean, alongside the numbers of critical-point births and deaths. 
All five curves share similar overall trends, indicating that MS-COOT captures much of the temporal variation identified by the baselines. Large distances frequently coincide with substantial critical-point changes, such as the ten births at $108\to109$, although this correspondence is not universal. During intervals without critical-point births or deaths, GWD often falls below 0.1 of its mean, whereas the other four profiles generally remain around 0.2 or higher. WD and FGW capture the continuous displacement of critical points through their spatial costs. MS-COOT also responds to vortex motion, despite not directly comparing critical-point coordinates across frames: as vortices move downstream while the near-wake region remains approximately stationary, their relative geometric relationships change within the MS complexes.

\para{Effects of probability weighting.}
The PI-weighted MS-COOT profile exhibits smaller fluctuations than WD, FGW, and uniform-weight MS-COOT over certain intervals, notably from $57\to58$ through $62\to63$. In this dataset, PI-based weights emphasize persistent critical points associated with vortices, increasing their relative contribution to the distance while reducing the influence of lower-weight features. This provides an interpretation of the smoother response to the continuous vortex evolution observed in these intervals.

\para{Sensitivity to region rearrangements.}
Both MS-COOT weighting settings also exhibit greater sensitivity than the baselines to certain region-level changes. To identify such cases, we compare the ratios of normalized MS-COOT and FGW distances across all 156 adjacent-step transitions. The transition $49\to50$ ranks among the five largest ratios and provides a representative example of this discrepancy (\cref{fig:vortexstreet}(c)). At this step, the normalized distances of WD, GWD, and FGW are 0.27, 0.17, and 0.26, respectively, compared with 0.98 and 1.30 for uniform- and PI-weighted MS-COOT. 
In the highlighted area, a saddle--maximum pair disappears, merging two neighboring regions and reducing the total region count from 89 to 88. Meanwhile, a minimum--saddle pair appears near the lower-left corner of the enlarged view without a comparable region-level rearrangement. These events offset each other in the critical-point count, which remains at 117. Although the baselines register relatively small distances, MS-COOT responds more strongly to the region merge through its augmented hypernetwork representation, which explicitly incorporates regions and their relationships to critical points.

Overall, MS-COOT captures the periodicity and major temporal changes identified by the baselines while providing additional sensitivity to geometric evolution and region-level rearrangements. The latter distinction becomes more pronounced in \cref{sec:ionization}.~\Cref{sec:appendix_timevarying} shows all eight configurations.

\begin{figure}[bt!]
\centering
\vspace{-2mm}
\includegraphics[width=0.96\columnwidth]{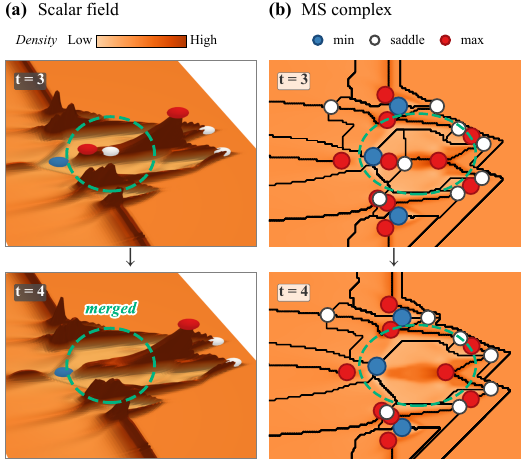}
\vspace{-2mm}
\caption{\textbf{Ionization Front~\cite{whalen2008ionization}: PI-weighted MS-COOT ranks the localized region merge at \(3\to4\) as the largest adjacent-step change, unlike the evaluated baselines.} (a)~Terrains of the density field at timesteps 3 and 4; the dashed circle marks the affected region, labeled merged at timestep 4. (b)~Corresponding MS complexes: from timestep 3 to 4 the separatrix inside the circle vanishes and the three regions it separated become one.}
\vspace{-2mm}
\label{fig:ionfront}
\end{figure}

\subsection{Ionization Front}
\label{sec:ionization}

The Ionization Front dataset originates from the 2008 IEEE SciVis Contest~\cite{whalen2008ionization,sciVis2008} and simulates a radiation-driven instability in primordial gas. An ionization front separating hot ionized gas from cold neutral gas propagates rightward, altering the underlying density field and its MS complex. Following Li et al.~\cite{li2023comparing}, we analyze the gas density on 2D slices near the center of the volume, selecting 30 consecutive timesteps for evaluation.

\para{Sensitivity to region merges.}
\Cref{fig:ionfront} highlights the transition $3\to4$, which produces the largest adjacent-step distance under PI-weighted MS-COOT among all 29 transitions, but does not rank first for any baseline or uniform-weight MS-COOT. The scalar fields (\cref{fig:ionfront}(a)) show the advancing ionization front, while the MS complexes (\cref{fig:ionfront}(b)) reveal a substantial region merge within the circled area. Between the two timesteps, a saddle--maximum pair disappears, removing a separatrix that previously divided the area into three regions. Despite this change, the surrounding critical-point configuration and graph geometry remain largely similar.

The baselines respond relatively weakly to this transition because the disappearance of only two critical points produces a limited change in their spatial distributions, while the shortest-path distances on the weighted 1-skeletons also does not change drastically. In contrast, MS-COOT explicitly incorporates regions and their relationships to critical points, capturing the substantial rearrangement caused by the merge. The PI-based measure further emphasizes the neighboring high-persistence maxima, while assigning less weight to the disappearing critical points. Together, the region-level structural change and the persistence-based weighting make this transition particularly prominent for PI-weighted MS-COOT.

Ablation results further illustrate these contributions. Replacing PI-based weights with uniform weights lowers the transition from first to 21st in the ranking of adjacent-step distances, indicating the contribution of persistence-based weighting to the pronounced response.~\Cref{sec:appendix_timevarying} provides the full distance matrices and adjacent-step profiles, while~\Cref{sec:appendix_cp_matching} examines the critical-point correspondences for this transition.

\begin{figure}[bt]
  \centering
  \includegraphics[width=0.95\linewidth]{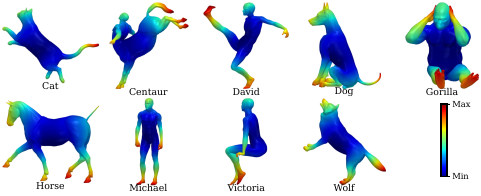}
  \vspace{-2mm}
  \caption{\textbf{TOSCA dataset: one representative mesh per category} colored by the average geodesic distance (AGD) scalar field, normalized to $[0,1]$. Extremities (paws, hands, tails) have high AGD (red); torso regions have low AGD (blue). The 9 categories span quadrupeds (cat, dog, horse, wolf), humanoids (David, Michael, Victoria), and mixed forms (centaur, gorilla).}
  \label{fig:tosca_gallery}
  \vspace{-3mm}
\end{figure}

\subsection{Shape Classification on TOSCA}
\label{sec:tosca}

We evaluate MS-COOT on the TOSCA non-rigid shape dataset~\cite{bronstein2008numerical}, which has also been used to evaluate merge-tree comparison methods~\cite{sridharamurthy2020edit,lyu2024fast}. We compare its shape-classification performance with three optimal-transport baselines and use the same dataset to examine the contributions of individual MS-COOT components.

\para{Data and preprocessing.}
TOSCA contains 80 meshes in nine categories (\cref{fig:tosca_gallery}): cat (11), centaur (6), David (7), dog (9), gorilla (4), horse (8), Michael (20), Victoria (12), and wolf (3). The meshes within each category represent different poses of a shape. Following Lyu et al.~\cite{lyu2024fast}, we construct an approximate average geodesic distance (AGD) field from 15 farthest-point-sampled anchor vertices, using the heat method~\cite{crane2017heat} to estimate geodesic distances. We normalize each field to $[0,1]$. AGD tends to be higher at limb extremities and lower near the torso (\cref{fig:tosca_gallery}), providing a deformation-tolerant scalar descriptor of the shape's intrinsic geometry.

We extract MS complexes using TTK~\cite{tierny2017topology} after persistence simplification at $1\%$ of each mesh's scalar range. This reduces the number of critical points (CPs) from 28--172 to 20--67 per mesh, leaving 9--34 regions. For gorilla meshes containing detached hair or fur components, we retain the largest connected component and recompute AGD before extracting the MS complex. All four methods use the resulting preprocessed meshes.

\para{Distance and evaluation settings.}
For TOSCA, we use a scalar-value sample cost in place of the default CP-type penalty of \cref{eq:type_penalty}:
\begin{equation}
  C[i,j] = \bigl|f(\mathrm{CP}_i)-g(\mathrm{CP}_j)\bigr|,
  \label{eq:scalar_cost}
\end{equation}
where $f$ and $g$ denote the normalized AGD fields of the two meshes. 
Because geodesic distances are preserved under intrinsic isometries, AGD is suited to comparing pose variations; the sampled and numerically approximated fields need not be exactly invariant. We use \cref{eq:scalar_cost} without an additional type penalty.

The MS-COOT reference configuration uses the augmented-graph shortest-path relation $\omega$ with centroid region representatives (\cref{sec:method}), the persistence-image node measure $\mu$ and its induced region measure $\nu$, and jointly optimized couplings $\pi$ and $\xi$. Its settings are $\alpha=0.5$, $\varepsilon=0.001$, and persistence-image bandwidth $\sigma=0.3$, with a $100\times100$ persistence-image grid. Unpaired CPs receive the default provisional weight $1/n$, followed by renormalization, as in \cref{sec:method}. The remaining solver settings follow \cref{sec:method} and~\Cref{sec:appendix_implementation}.

We use the same scalar cost for WD and the feature term of FGW. GWD compares shortest-path distances on the Euclidean-weighted CP--CP graph, normalized separately for each mesh; FGW combines this graph-based structural cost with the scalar feature cost. The baselines use unregularized OT solvers by default. For classification, we compute pairwise distances between all 80 meshes and perform leave-one-out $k$-nearest-neighbor ($k$-NN) evaluation. We report $k=1$ results, selecting each method's configuration by the highest leave-one-out $k=1$ accuracy and breaking ties using $k=3$ accuracy. The evaluated grids contain 96, 32, 864, and 18 configurations for WD, GWD, FGW, and MS-COOT, respectively. 
Since the same dataset is used for configuration selection and evaluation, we consider these as the best observed leave-one-out results instead of independent test-set estimates.

\begin{table}[bt]
\centering
\small
\setlength{\tabcolsep}{4pt}
\caption{\textbf{TOSCA shape classification.} Per-class recall under leave-one-out $1$-NN evaluation, using the selected configuration for each method. $N$ is the number of meshes in each class; bold indicates the highest recall in a row. Overall recall is micro-averaged across all 80 meshes.}
\label{tab:tosca}
\begin{tabular*}{\columnwidth}{@{\extracolsep{\fill}}lcccc@{}}
\toprule
Class ($N$) & WD & GWD & FGW & MS-COOT \\
\midrule
Cat (11)      & \textbf{1.0000} & 0.8182 & \textbf{1.0000} & \textbf{1.0000} \\
Centaur (6)   & \textbf{1.0000} & \textbf{1.0000} & \textbf{1.0000} & \textbf{1.0000} \\
David (7)     & \textbf{0.5714} & 0.0000 & 0.2857 & \textbf{0.5714} \\
Dog (9)       & 0.6667 & 0.6667 & \textbf{0.7778} & \textbf{0.7778} \\
Gorilla (4)   & \textbf{1.0000} & 0.2500 & \textbf{1.0000} & \textbf{1.0000} \\
Horse (8)     & 0.8750 & 0.6250 & \textbf{1.0000} & \textbf{1.0000} \\
Michael (20)  & \textbf{0.8000} & 0.7500 & \textbf{0.8000} & 0.7500 \\
Victoria (12) & 0.9167 & 0.7500 & 0.9167 & \textbf{1.0000} \\
Wolf (3)      & 0.6667 & \textbf{1.0000} & \textbf{1.0000} & \textbf{1.0000} \\[2pt]
\midrule
\textbf{Overall (80)} & 0.8375 & 0.6750 & 0.8500 & \textbf{0.8750} \\
\bottomrule
\end{tabular*}
\vspace{-3mm}
\end{table}

\para{Classification results.}
\Cref{tab:tosca} shows that MS-COOT correctly classifies 70 of 80 meshes (0.8750), compared with 68 for FGW (0.8500), 67 for WD (0.8375), and 54 for GWD (0.6750). 
MS-COOT has the highest observed overall recall and individual ones in eight out of nine classes across all tested methods, and achieves perfect recall in six of nine classes. 
Its ten errors include three David meshes and five Michael meshes; these two humanoid categories have similar AGD-derived structures.
The result demonstrates competitive classification performance while retaining the explicit region correspondence unavailable from the three CP-based baselines.
Further analysis, including the confusion analysis and case studies for intra-class and inter-class correspondences, are reported in~\Cref{sec:appendix_tosca}.

\subsection{Ablation on TOSCA}
\label{sec:ablation}

We evaluate the sensitivity of MS-COOT to its hypernetwork function, probability measures, and parameter choices using the TOSCA classification task. Each ablation changes one component of the reference configuration while retaining the scalar sample cost of \cref{eq:scalar_cost}, $\alpha=0.5$, $\sigma=0.3$, and the normalization and solver settings described above, unless otherwise specified. We report results under both the default entropic regularization ($\varepsilon=0.001$) and unregularized OT subproblems ($\varepsilon=0$). The latter removes entropy from each transport update but does not guarantee a globally optimal solution to the joint nonconvex problem.

For the hypernetwork function $\omega$, we compare weighted shortest-path distances on the augmented graph against three alternatives: (i) hop-based proximity, which assigns unit weight to every CP--CP, CP--RC, and RC--RC edge of the same augmented graph before computing shortest-path distances; (ii) direct Euclidean distances between CPs and region centroids; and (iii) binary CP--region incidence, equal to one when a CP bounds a region and zero otherwise. Each relation is normalized per complex as in \cref{sec:method}.
For the node measure $\mu$, we compare persistence-image weights with uniform and raw-persistence weights, recomputing the induced region measure $\nu$ accordingly. 
We also examine the treatment of CPs without finite persistence pairs by replacing their default provisional weight $1/n$ with either $0.1/n$ or the maximum unnormalized persistence-image weight among CPs with finite persistence pairs. In each case, the resulting weights are normalized jointly over all CPs, and the induced region measure $\nu$ is recomputed accordingly.
\Cref{tab:ablation} reports leave-one-out $1$-NN recall over 80 meshes, where each correct classification contributes $1/80=0.0125$ to the overall recall.

\begin{table}[bt]
\centering
\scriptsize
\setlength{\tabcolsep}{1.8pt}
\caption{\textbf{MS-COOT component ablation on TOSCA.} Leave-one-out $1$-NN recall over 80 meshes. Each variant changes one component from the reference configuration; $\Delta$ denotes the difference in recall at the same $\varepsilon$. 
Binary incidence is reported only at $\varepsilon=0$, as MS-COOT's BCD optimization failed to converge on 29\% of mesh pairs at $\varepsilon=0.001$.}
\label{tab:ablation}
\begin{tabular}{@{}llcccc@{}}
\toprule
& & \multicolumn{2}{c}{$\varepsilon=0.001$} & \multicolumn{2}{c}{$\varepsilon=0$} \\
\cmidrule(lr){3-4} \cmidrule(lr){5-6}
Component & Variant & Recall & $\Delta$ & Recall & $\Delta$ \\
\midrule
\multicolumn{2}{@{}l}{Reference} & 0.8750 & & 0.8375 & \\
\midrule
$\omega$ & hop distance & 0.7125 & $-0.1625$ & 0.6875 & $-0.1500$ \\
$\omega$ & Euclidean CP--RC & 0.8125 & $-0.0625$ & 0.8250 & $-0.0125$ \\
$\omega$ & binary incidence & --- & --- & 0.8625 & $+0.0250$ \\
\midrule
$\mu$ & uniform & 0.8500 & $-0.0250$ & 0.8250 & $-0.0125$ \\
$\mu$ & raw persistence & 0.8000 & $-0.0750$ & 0.7625 & $-0.0750$ \\
\midrule
Unpaired CP & $0.1/n$ weight & 0.9000 & $+0.0250$ & 0.8750 & $+0.0375$ \\
Unpaired CP & max finite weight & 0.8750 & $0.0000$ & 0.8625 & $+0.0250$ \\
\bottomrule
\end{tabular}
\vspace{-3mm}
\end{table}

\para{Hypernetwork function.}
Replacing weighted shortest-path distances with hop-based proximity reduces recall by 13 meshes at $\varepsilon=0.001$ and 12 at $\varepsilon=0$. This indicates that incorporating geometric edge lengths into the relational representation is beneficial relative to unweighted hop proximity on this task. Direct Euclidean CP--centroid distances produce smaller reductions of five and one mesh, respectively. 
Binary incidence achieves a recall of 0.8625 at $\varepsilon=0$, exceeding the shortest-path reference (0.8375) by two meshes. However, with binary incidence at $\varepsilon=0.001$, the BCD optimization fails to converge on 29\% of mesh pairs, preventing a complete classification evaluation under the default regularization setting. 
In summary, the shortest-path construction provides a graded representation of CP--region relationships that incorporates both connectivity and geometric separation, while converging under the tested default configuration.

\para{Probability measures.}
Uniform node weights yield recalls of 0.8500 and 0.8250 at $\varepsilon=0.001$ and $0$, respectively, corresponding to only two and one fewer correct classifications than the persistence-image reference. In contrast, raw-persistence weighting reduces recall by six meshes at both regularization levels. 
On TOSCA, persistence-image weighting improves classification relative to raw persistence, but offers only a small observed advantage over uniform weights. 

\para{Unpaired critical points.}
Changing the provisional weights of CPs without finite persistence pairs changes the number of correct classifications by at most three meshes relative to the reference. Assigning $0.1/n$ produces the highest recall among the tested variants, reaching 0.9000 at $\varepsilon=0.001$ and 0.8750 at $\varepsilon=0$. However, these results do not establish that this choice generalizes better than the default $1/n$. We retain the default treatment and leave a broader assessment of unpaired-CP weighting across different mesh types for future work.

\begin{figure}[bt]
\centering
\includegraphics[width=0.9\linewidth]{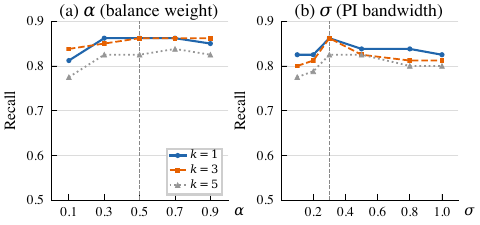}
\vspace{-2mm}
\caption{\textbf{MS-COOT parameter sensitivity on TOSCA.} Leave-one-out $k$-NN overall recall for $k\in\{1,3,5\}$ while varying one parameter from the reference configuration: (a) sample-cost weight $\alpha$ (default $0.5$) and (b) persistence-image bandwidth $\sigma$ (default $0.3$). Dashed lines indicate the defaults; all other settings are fixed.}
\vspace{-3mm}
\label{fig:tosca_sweeps}
\end{figure}

\para{Parameter sensitivity.}
\Cref{fig:tosca_sweeps} shows relatively similar recall across $\alpha\in[0.3,0.9]$, with lower recall at $\alpha=0.1$, where the sample-cost term contributes less to the objective. Among the tested bandwidths, $\sigma=0.3$ achieves the highest $k=1$ recall, with modest reductions toward both ends of the sweep. This behavior is consistent with the limiting effects of persistence-image smoothing: narrow kernels place greater emphasis on individual persistence values, whereas broad kernels tend toward more uniform weights. These results characterize sensitivity around the selected configuration rather than establishing parameter robustness beyond the tested ranges.

Overall, the ablation indicates that the choice of hypernetwork function and probability measure can affect MS-COOT's classification performance. Weighted shortest-path proximity substantially improves recall over hop-based proximity, while binary incidence remains competitive under unregularized transport. Persistence-image weighting improves upon raw persistence but yields results close to uniform weighting. Together, these observations support the selected configuration as a practical choice for the evaluated applications, without implying that its individual components are uniquely optimal for shape classification.

\subsection{Viscous Finger}
\label{sec:viscous}

To evaluate MS-COOT on three-dimensional volumetric data, we use the Viscous Finger ensemble from the 2016 IEEE SciVis Contest~\cite{sciVis2016}, previously studied using topology-based feature tracking~\cite{lukasczyk2017viscous}. The dataset consists of finite pointset method (FPM) simulations of salt dissolving in water, where instabilities produce finger-like structures with elevated salt concentration. We examine region correspondence across timesteps and the ability of MS-COOT to distinguish simulation runs generated under different smoothing-length settings.

For the temporal correspondence example, we use the first run at smoothing length $h=0.30$, interpolating its particle concentration data onto a $50\times50\times50$ grid over 120 timesteps. After persistence simplification at 6\% of the global scalar range, the extracted MS complexes contain 1--17 regions per timestep. Each simplified complex has a single global minimum, with regions corresponding to the descending manifolds of local maxima. The high-concentration portions of these regions capture structures associated with the viscous fingers. For visualization in \cref{fig:teaser}, we display only concentrations above 30 (from a range extending to approximately $3.2\times10^4$) to suppress the low-concentration background. This threshold affects the visualization rather than the MS complex extraction.

\para{Region correspondence.}
\Cref{fig:teaser}(a,b) shows two consecutive MS complexes at $t=51$ and $t=52$, containing nine and five regions, respectively. The region coupling $\xi$ (\cref{fig:teaser}(c)) identifies correspondences consistent with the spatial organization of these structures. For example, S1 and S3 both match T1, while S5, S6, and S8 match T4, suggesting merging of the corresponding regions. Other regions retain dominant correspondences, including S4 $\rightarrow$ T2 and S7 $\rightarrow$ T3. Regions S2 and S9 distribute their mass across multiple targets, indicating splitting and reorganization into multiple target regions.

These assignments illustrate how $\xi$ relates segmented structures despite changes in region count. Unlike the CP-level couplings produced by the evaluated WD, GWD, and FGW baselines, $\xi$ directly provides a region-indexed transport plan, enabling analysis of region-level changes.

\para{Smoothing-length discrimination.}
The ensemble contains simulations at three smoothing lengths of the particle method ($h=0.20$, $0.30$, and $0.44$). The smoothing length controls the particle interaction radius and affects the resulting viscous-fingering dynamics. 
Previous analyses of this ensemble reported systematic
differences across simulation settings in finger counts,
onset times, propagation rates, and aggregate structural
statistics~\cite{lukasczyk2017viscous,GralkaGrottel2018}.
These observations motivate examining whether such
differences are reflected in distances between MS complexes.
All scalar field instances undergo the same preprocessing steps as those used in the temporal correspondence example above.

We select five independent runs per smoothing-length setting, yielding 15 runs in total. For each pair of runs, we average the distances computed at six timestep indices ($t\in\{60,70,80,90,100,110\}$). We focus on later timesteps, when the viscous fingers have developed. All methods use fixed parameters without tuning to the smoothing-length labels.
The evaluated WD, GWD, and FGW baselines use uniform CP weights; for MS-COOT, we compare uniform and persistence-image (PI) weights while retaining the remaining settings. 

The multidimensional scaling (MDS) embeddings in \cref{fig:teaser}(d) show clearer grouping by smoothing length for PI-weighted MS-COOT than for the evaluated baselines. To quantify discrimination independently of the two-dimensional embeddings, we perform leave-one-out 1-nearest-neighbor (1-NN) classification using the original pairwise distances. As shown in \cref{tab:vf_classification}, PI-weighted MS-COOT achieves 80.0\% accuracy (12/15 correct), compared with 53.3\% for WD, 46.7\% for GWD, and 33.3\% for FGW. A permutation test with 1{,}000 class-count-preserving label shuffles yields $p=0.002$ for PI-weighted MS-COOT, indicating an association between the computed distances and smoothing-length labels under the tested null hypothesis. 


The probability measure has a substantial effect on MS-COOT in this experiment. With uniform weights, classification accuracy decreases to 46.7\% (7/15 correct), compared with 80.0\% (12/15) under PI-based weighting. Notably, the MS-COOT sample cost only checks consistency of CP types between critical points. The observed discrimination therefore relies on the CP--region relational representation and the associated probability measures. The improvement under PI-based weighting is consistent with emphasizing informative CP--region relationships, although the experiment does not isolate which relationships account for the difference.

\para{Interpretation.}
The Viscous Finger results suggest a setting in which CP--region relationships may offer advantages over CP-only graph representations. 
Across independent simulation runs, local critical points and separatrices can vary even when the broader organization of the finger-like structures remains similar. 
In the augmented graph, region centers connect multiple boundary CPs and adjacent regions, providing alternative shortest paths that may reduce the sensitivity of CP--region distances to the disappearance of individual separatrices. 
This may be particularly relevant here, where the MS complexes contain approximately 460 CPs but only 10 regions on average (see~\cref{tab:implementation}). 
The higher classification accuracy under PI-based weighting is consistent with persistence-based measures further emphasizing informative relationships within this representation. 
Although this mechanism is not isolated experimentally, it offers a possible explanation for the observed discrimination of smoothing-length settings.

\para{Summary.}
The Viscous Finger experiment demonstrates two benefits of MS-COOT on volumetric simulation data. 
First, the region coupling $\xi$ provides explicit correspondences between segmented flow structures, revealing region consolidation and partial matches across timesteps.
Second, PI-weighted MS-COOT achieves the highest classification accuracy (80.0\%) among the evaluated configurations, substantially improving upon its uniform-weight variant (46.7\%) and the CP-based baselines. 
These findings illustrate the potential of combining region-level relational geometry with persistence-based weighting to distinguish independent simulations despite variations in their detailed critical-point structures.

\begin{table}[bt]
\centering
\caption{\textbf{Viscous Finger smoothing-length classification.} Leave-one-out 1-NN accuracy over 15 simulation runs (five per smoothing-length setting). Per-class recall is reported as correctly classified runs out of five. Permutation-test $p$-values use 1{,}000 label shuffles: $^{*}$\,$p{<}0.05$; $^{**}$\,$p{<}0.01$. MS-COOT is reported with uniform and persistence-image (PI) weights; \cref{fig:teaser}(d) shows the PI setting.}
\vspace{-2mm}
\label{tab:vf_classification}
\small
\resizebox{0.98\columnwidth}{!}{%
\begin{tabular}{lccccc}
\toprule
 & & & \multicolumn{3}{c}{Per-class recall}\\
\cmidrule(lr){4-6}
Method & Accuracy & $p$-value & Coarse & Medium & Fine \\
\midrule
WD     & 53.3\% & 0.028$^{*}$ & 0/5 & 5/5 & 3/5 \\
GWD    & 46.7\% & 0.160       & 3/5 & 3/5 & 1/5 \\
FGW    & 33.3\% & 0.427       & 3/5 & 1/5 & 1/5 \\
MS-COOT (uniform) & 46.7\% & 0.133 & 0/5 & 5/5 & 2/5 \\
\textbf{MS-COOT (PI) } & \textbf{80.0\%} & \textbf{0.002$^{**}$} & 3/5 & 5/5 & 4/5 \\
\bottomrule
\end{tabular}%
}
\vspace{-2mm}
\end{table}

\subsection{Runtime} 
\Cref{tab:implementation} reports per-pair times for every method and dataset. A single MS-COOT distance takes between $0.5$ and $6$\,s on the 2D datasets and on TOSCA, whose simplified complexes average $34$ to $117$ critical points, and about a minute on Viscous Finger, where they average $467$. FGW is about three orders of magnitude faster per pair; WD, which solves one linear program with no structure term, is faster still. 
The difference is structural: FGW takes a few conditional-gradient steps, each an exact linear transport problem, while MS-COOT alternates two entropic couplings over several iterations, from several starting points. 

We note that the Sinkhorn algorithm is well suited to parallel computation, as its iterations primarily consist of matrix–vector products and elementwise scaling operations that can be performed concurrently~\cite{cuturi2013sinkhorn,Schmitzer2019}. We leave the extension to scalable computation for future work.

\begin{table}[hbt]
\centering
\caption{Dataset statistics and per-pair runtime in ms, mean over 100 pairs per dataset; $\pm$ is the standard deviation over five repetitions of that mean. $N$: instances (timesteps, or meshes for TOSCA). Reg.: regions. VS: Vortex Street, IF: Ionization Front, VF: Viscous Finger.}
\label{tab:implementation}
\footnotesize
\setlength{\tabcolsep}{2.5pt}
\renewcommand{\arraystretch}{0.95}
\begin{tabular}{lcc cccc}
\toprule
 & Avg & Avg & \multicolumn{4}{c}{Runtime (ms/pair)} \\
\cmidrule(lr){4-7}
Dataset & CPs & Reg. & WD & GWD & FGW & MS-COOT \\
\midrule
VS    & 117 & 88  & $0.51{\pm}0.02$ & $8.84{\pm}0.12$ & $3.74{\pm}0.04$ & $5937{\pm}108$ \\
IF    & 34  & 29  & $0.34{\pm}0.05$ & $1.27{\pm}0.37$ & $0.36{\pm}0.03$ & $546{\pm}2$ \\
TOSCA & 43  & 21  & $0.49{\pm}0.06$ & $2.25{\pm}0.52$ & $0.61{\pm}0.02$ & $563{\pm}4$ \\
VF    & 467 & 10  & $8.97{\pm}0.14$ & $1017{\pm}25$   & $162{\pm}3$     & $54178{\pm}1401$ \\
\bottomrule
\end{tabular}
\end{table}

\section{Conclusion and Discussion}
\label{sec:conclusion}

We presented MS-COOT, a co-optimal transport distance for Morse--Smale complexes that jointly computes critical-point and region correspondences. By representing critical points and regions as nodes and hyperedges of a measure hypernetwork, MS-COOT incorporates both geometric and structural relationships into a unified comparison framework. 
Compared to graph-based OT distances, MS-COOT produces an explicit region coupling $\xi$, enabling the analysis of region splits, merges, and partial correspondences.

Our experiments characterize MS-COOT's sensitivity to changes in MS complex structure and feature significance.
The controlled examples demonstrate that the distance responds to changes in CP--region geometry, while persistence-image (PI) weighting helps distinguish low-persistence perturbations from more prominent features and mitigates mismatched transport caused by changes in persistence pairing. 
On time-varying simulations, MS-COOT captures periodic behavior and responds to region rearrangements that produce relatively small changes in graph-based distances. 
On TOSCA, it achieves competitive shape classification accuracy while retaining region-level correspondences. 
On Viscous Finger, it identifies region merges and partial matches in volumetric data and provides improved discrimination of particle smoothing-length settings.

\para{Parameter choices.}
MS-COOT provides a flexible framework in which the CP--region relation, probability measures, and sample costs can be adapted to different applications.
Our TOSCA ablation demonstrates the importance of these choices without establishing a uniquely optimal configuration.
Weighted shortest-path proximity outperforms hop-based proximity, supporting the use of geometric distances alongside connectivity.
PI-based weighting improves classification over raw persistence but performs similarly to uniform weighting.
However, its dependence on the local distribution of persistence pairs may not always reflect task-relevant feature importance, making simpler persistence transforms viable alternatives.
For CPs without finite persistence pairs, our default provisional weight of $1/n$ avoids assuming unusually high or low significance, yet alternative weights can be beneficial.

More generally, structural and geometric information is useful when it captures meaningful differences, but may introduce undesirable sensitivity when not preserved under the variations of interest.
The appropriate choices are therefore task-dependent: uniform and PI-based weights yield similar accuracy on TOSCA, whereas PI weighting substantially improves classification on Viscous Finger.
Normalization is another consideration: our per-complex maximum normalization may introduce sensitivity to path-length distributions without corresponding topological changes; dataset-wide normalization is a possible alternative whose effects remain untested.
Overall, our default configuration combines shortest-path CP--region relations, centroid region representatives, and PI-based measures, providing a practical starting point that performs well across the evaluated datasets and can be adapted to application-specific invariances and features.

\para{Metric properties.}
When $\alpha=0$, the square root of Eq.~(1) inherits the pseudometric properties of COOT with squared ground loss~\cite{redko2020co}; the reported distance is the squared quantity. 
For $\alpha>0$, non-negativity and symmetry are preserved for the sample costs considered here, while the triangle inequality remains unestablished. 

\para{Limitations.}
MS-COOT has several limitations. First, its bilinear objective is non-convex, and block coordinate descent does not guarantee global optimality. Second, jointly optimizing two transport couplings incurs substantially higher computational cost than graph-based OT baselines, with per-pair runtimes typically two to three orders of magnitude higher than FGW in our experiments. 
However, the Sinkhorn updates are amenable to parallel computation, offering opportunities to accelerate the solver, although the sequential alternation between couplings remains a challenge. 
Third, MS-COOT depends on persistence simplification, whose threshold affects the extracted regions and their correspondences. 
Finally, the two couplings are not explicitly constrained to agree on region boundaries, and region correspondences may be ambiguous in symmetric configurations.

\para{Future directions.}
A natural extension is to develop region-level feature tracking based on $\xi$, incorporating geometric and temporal consistency to resolve ambiguous correspondences. 
Beyond tracking, MS-COOT could support machine learning tasks that benefit from structural comparisons of scalar fields, such as classification for medical images.
Improving computational scalability through parallel optimization is a further priority. 
On the theoretical side, establishing stability and metric properties for the augmented formulation would clarify its behavior under perturbations of scalar fields.

\ifCLASSOPTIONcompsoc
  \section*{Acknowledgments}
\else
  \section*{Acknowledgment}
\fi
This project was partially supported by NSF under grant CCF-2444309.
Mingzhe Li is supported by the Provost’s Postdoctoral Fellowship at the University of Notre Dame.
ChatGPT-5.6 Sol was used for language polishing.

\bibliographystyle{IEEEtran}
\bibliography{VIS'26-MSCOOT}

\vskip -2.2\baselineskip plus -1fil
\vspace{-2mm}
\begin{IEEEbiography}[\vspace{-6mm}{\includegraphics[width=1in,height=1.25in,clip,keepaspectratio]{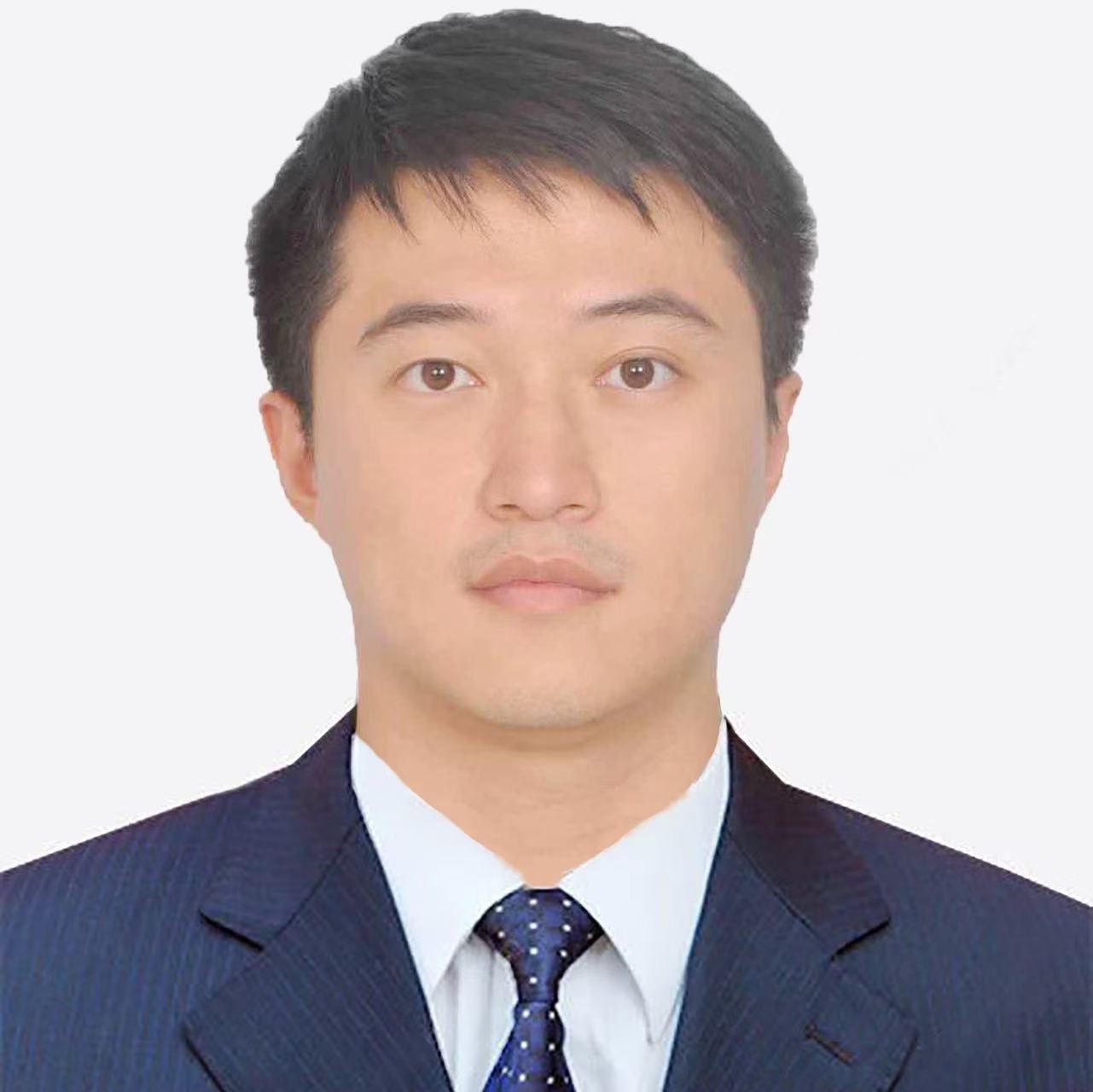}}]{Guangyu Meng} is a Ph.D. student in the Department of Computer Science and Engineering at the University of Notre Dame. He received his B.S. degree in Electronic Science and Engineering from Huazhong University of Science and Technology, Hubei, China. His research interests lie at the intersection of topological data analysis, machine learning, computational geometry, and medical image analysis.
\end{IEEEbiography}
\vskip -2.2\baselineskip plus -1fil
\vspace{-4mm}
\begin{IEEEbiography}[\vspace{-6mm}{\includegraphics[width=1in,height=1.25in,clip,keepaspectratio]{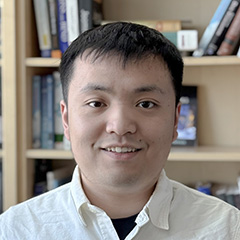}}]{Mingzhe Li} is a postdoctoral research fellow in the Department of Computer Science and Engineering at the University of Notre Dame. He received his PhD in Computing (Data Management and Analysis) at the University of Utah in 2025. His research interests include topological data analysis, scientific visualization, uncertainty visualization, optimal transport, high-performance computing, and machine learning.
\end{IEEEbiography}
\vskip -2.2\baselineskip plus -1fil
\vspace{-4mm}
\begin{IEEEbiography}[\vspace{-6mm}{\includegraphics[width=1in,height=1.25in,clip,keepaspectratio]{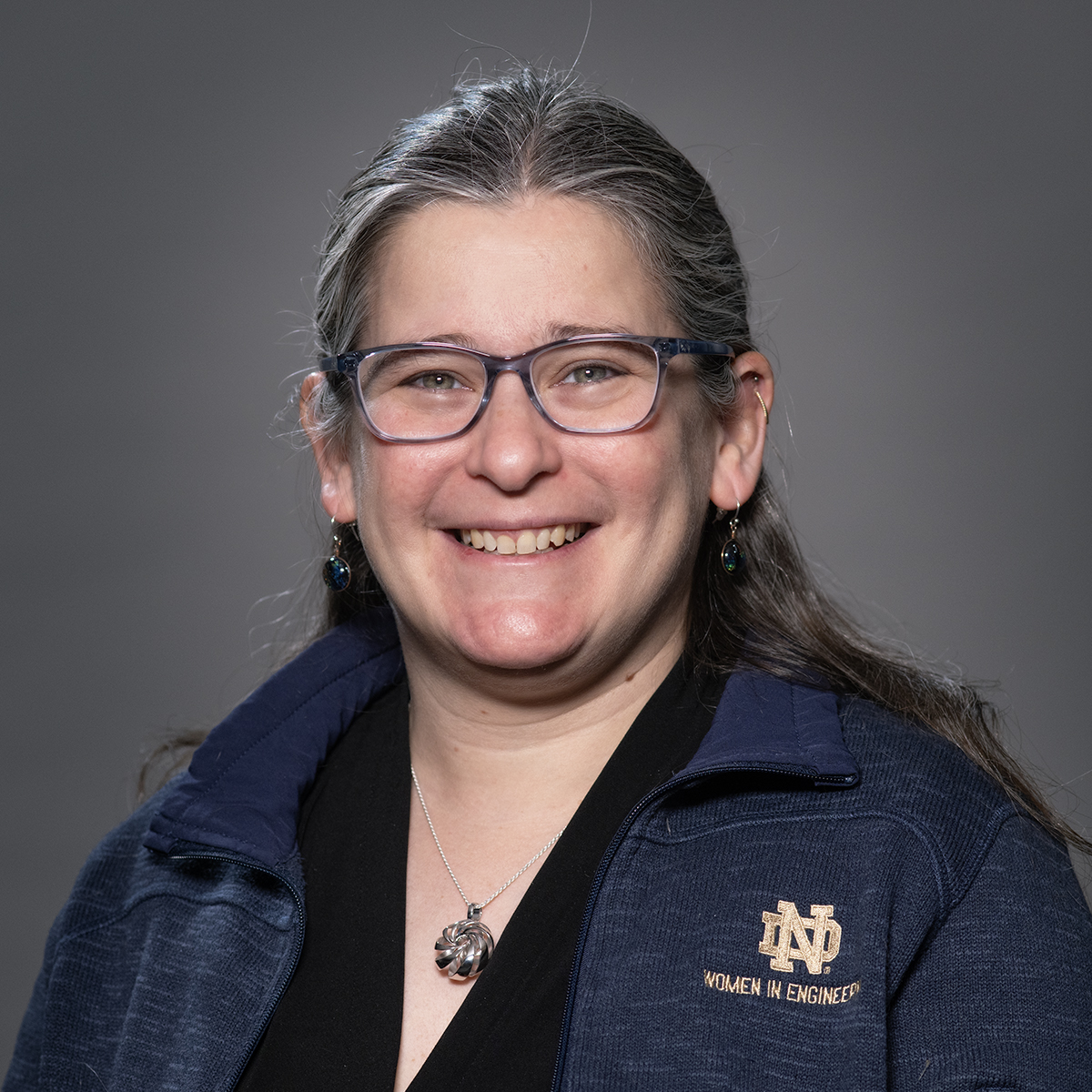}}]{Erin W. Chambers}  is the Schnyder Mission Family Collegiate Professor in the Department of Computer Science and Engineering at the University of Notre Dame, with a concurrent appointment in the Department of Applied and Computational Mathematics and Statistics.  She received her PhD in Computer Science in 2008 and her MS in Mathematics in 2006 from the University of Illinois at Urbana-Champaign.  Her research focus is on computational topology and geometry, with a more general interest in combinatorics and combinatorial algorithms.
\end{IEEEbiography}

\ifincludesupplement
    \clearpage
    \appendices
    \crefalias{section}{appendix}
    \crefname{appendix}{Appendix}{Appendices}
    \Crefname{appendix}{Appendix}{Appendices}
\providecommand{\FloatBarrier}{}
\renewcommand{\topfraction}{0.95}
\renewcommand{\bottomfraction}{0.6}
\renewcommand{\textfraction}{0.05}
\renewcommand{\floatpagefraction}{0.7}
\renewcommand{\dbltopfraction}{0.95}
\renewcommand{\dblfloatpagefraction}{0.7}
\setcounter{topnumber}{4}
\setcounter{bottomnumber}{2}
\setcounter{totalnumber}{6}
\setcounter{dbltopnumber}{3}

This document provides supplementary material, including notation (\cref{sec:appendix_notation}), implementation details (\cref{sec:appendix_implementation}), additional TOSCA results (\cref{sec:appendix_tosca}), a comparison of node measures (\cref{sec:appendix_measures}), 
additional results on the time-varying datasets, including the Heated Cylinder experiment (\cref{sec:appendix_timevarying}), a measurement of how often the two couplings disagree (\cref{sec:appendix_consistency}), and a comparison of the critical point couplings for all four methods (\cref{sec:appendix_cp_matching}).

\section{Notation}
\label{sec:appendix_notation}
\Cref{tab:notation} summarizes the notation used throughout the main paper.

\begin{table}[t]
\centering
\caption{Summary of notation.}
\label{tab:notation}
\small
\setlength{\tabcolsep}{4pt}
\renewcommand{\arraystretch}{1}
\begin{tabular}{@{}lp{5.2cm}@{}}
\toprule
\textbf{Symbol} & \textbf{Description} \\
\midrule
\multicolumn{2}{l}{\textit{Input and Morse-Smale Complex}} \\
$f, g$ & Scalar fields \\
$\mathrm{CP}_i$ & The $i$-th critical point \\
$R_j$ & The $j$-th region (2-cell in 2-manifolds, 3-cell in 3-manifolds) \\
$\partial R_j$ & Set of boundary CPs of region $R_j$ \\
$n, m$ & Number of CPs and regions \\
$n_f, n_g, m_f, m_g$ & Number of CPs and regions for $f$ and $g$ \\
$\mathrm{type}(\cdot)$ & CP type: 0\,(min), 1\,(saddle), 2\,(max) in 2D; 0/1/2/3 in 3D \\
$b_i,\, d_i$ & Birth and death coordinates of $\mathrm{CP}_i$'s persistence pair \\
$p_i = \mathrm{pers}(\mathrm{CP}_i)$ & Persistence: $d_i - b_i$ \\
\midrule
\multicolumn{2}{l}{\textit{Hypernetwork and Augmented Graph}} \\
$\mathcal{H}_f, \mathcal{H}_g$ & Measure hypernetworks $(V, E, \mu, \nu, \omega)$ \\
$V,\; E$ & Node set (CPs) and hyperedge set (regions) \\
$G = (V_G, E_G)$ & Augmented graph; $V_G = V \cup \{\mathrm{RC}_j\}$ \\
$\mathrm{RC}_j$ & Region center of $R_j$: centroid of the spatial samples assigned to \(R_j\) \\
$d_G(\cdot,\cdot)$ & Shortest-path distance on $G$ \\
$\omega \in \mathbb{R}^{n \times m}$ & Hypernetwork function; $\omega[i,j] = d_G(\mathrm{CP}_i, \mathrm{RC}_j)$ \\
$Z_f, Z_g$ & Per-complex normalization factors: $Z_f = \max_{i,k} \omega_f[i,k]$, $Z_g = \max_{j,l} \omega_g[j,l]$ \\
\midrule
\multicolumn{2}{l}{\textit{Measures}} \\
$\mu \in \mathbb{R}^{n}$ & Node measure on CPs (persistence-based) \\
$\nu \in \mathbb{R}^{m}$ & Hyperedge measure on regions (from $\mu$) \\
$\sigma$ & PI Gaussian kernel bandwidth \\
\midrule
\multicolumn{2}{l}{\textit{MS-COOT Formulation}} \\
$\pi \in \mathbb{R}^{n_f \times n_g}$ & CP coupling \\
$\xi \in \mathbb{R}^{m_f \times m_g}$ & Region coupling \\
$\Pi(\mu_f, \mu_g)$ & Set of couplings with marginals $\mu_f, \mu_g$ \\
$C \in \mathbb{R}^{n_f \times n_g}$ & Sample cost (type penalty or scalar difference) \\
$\alpha$ & Balance weight (structure vs.\ sample cost) \\
$\varepsilon$ & Sinkhorn regularization parameter \\
\bottomrule
\end{tabular}
\end{table}

\FloatBarrier

\section{Implementation Details}
\label{sec:appendix_implementation}

This section records the information needed to reproduce the experiments: the hardware, the parameters shared by every dataset, the solver's starting couplings and stopping rule, the discretization of the persistence image, the persistence thresholds and the curves that justify them, and the costs and solvers of the baselines. Dataset-specific choices are stated here once, and the main paper refers to this section for them. The TOSCA experiments were run single-threaded on nodes of the Notre Dame Center for Research Computing; all other experiments were run on an Intel i9-12900K CPU with 128\,GB RAM.

\para{Parameter settings.} \Cref{tab:params} summarizes the MS-COOT parameters, which are shared across all five datasets; the dataset-specific settings are the persistence simplification threshold (\cref{tab:persistence}) and the sample cost. For the time-varying and Viscous Finger datasets, the critical-point sample cost (\Cref{eq:type_penalty}) is the type cost by default; TOSCA uses the scalar cost of~\cref{eq:scalar_cost} alone.
The Sinkhorn regularization $\varepsilon = 0.001$ follows standard practice in optimal transport~\cite{cuturi2013sinkhorn}, balancing numerical stability and approximation quality. We set the balance weight $\alpha = 0.5$. The persistence-image bandwidth $\sigma = 0.3$ sets how far neighboring persistence pairs contribute to a critical point's weight (\Cref{sec:measures} of the main paper). Sensitivity to $\alpha$ and $\sigma$ is shown in~\cref{fig:tosca_sweeps} of the main paper, and $\varepsilon = 0.001$ is compared against exact transport in~\Cref{tab:ablation} of the main paper. Because these parameters are stable on TOSCA, the dataset with the most diverse MS complexes, we use the same values of $\varepsilon$, $\alpha$, and $\sigma$ on every other dataset without further tuning.

\begin{table}[t]
\centering
\caption{Parameter settings for MS-COOT.}
\label{tab:params}
\small
\begin{tabular}{@{}lr@{}}
\toprule
\textbf{Parameter} & \textbf{Value} \\
\midrule
Sinkhorn regularization ($\varepsilon$) & 0.001 \\
Sample cost weight ($\alpha$) & 0.5 \\
PI bandwidth ($\sigma$) & 0.3 \\
PI grid resolution & $100 \times 100$ \\
Max BCD iterations & 200 \\
Max Sinkhorn iterations & 200 \\
Objective tolerance (BCD) & $10^{-6}$\\
Coupling tolerance ($\ell_1$ on $\pi$) & $10^{-7}$\\
Sinkhorn tolerance & $10^{-7}$\\
Sinkhorn variant & log-domain \\
\bottomrule
\end{tabular}
\end{table}

\para{Solver.} MS-COOT alternates Sinkhorn updates of $\pi$ and $\xi$ (\Cref{sec:optimization} of the main paper). Each iteration first updates $\pi$ from the current $\xi$ and then $\xi$ from the new $\pi$, so a run is determined by its starting region coupling. We run the alternation from 21 starting couplings: one deterministic start (the product coupling $\nu_f \nu_g^{\top}$) and 20 random couplings in $\Pi(\nu_f, \nu_g)$, each obtained by projecting a random positive matrix onto the marginals. We fix the random seed for reproducibility and retain the solution with the lowest value of the objective in~\cref{eq:mscoot} of the main paper, excluding the entropic terms used to regularize each subproblem. The random starts are not a formality: for the Vortex Street matrices, one of them attains the lowest objective for 99.6\% of pairs. On TOSCA, the deterministic start is instead the entropic transport plan of the sample cost averaged over the boundary critical points of each pair of regions, with regularization 0.05. A run stops when the $\ell_1$ change in $\pi$ between iterations falls below $10^{-7}$ or the absolute change in the entropically regularized objective falls below $10^{-6}$, with at most 200 iterations; each Sinkhorn solve runs in the log domain for at most 200 iterations with a tolerance of $10^{-7}$. If Sinkhorn fails to return a finite coupling, the update is re-solved exactly at $\varepsilon = 0$ and the alternation continues; this fallback was triggered on 5 of the 435 Ionization Front pairs and on no other pairs in the time-varying datasets.

\para{Persistence image.} Scalar fields are normalized to $[0,1]$ before persistence is computed, so $\sigma$ is in those units. The persistence surface of~\cref{eq:surface} of the main paper is discretized on a regular $100 \times 100$ grid spanning the bounding box of the diagram expanded by $3\sigma$ in each direction, with the persistence axis clipped at zero; each Gaussian is evaluated analytically at every grid center, with no interpolation or lookup, and the products are summed. The sum omits the area of the grid bins, so $w_i$ is proportional to, rather than equal to, the inner product; the constant cancels during normalization. At this resolution, the weights agree with the closed form to within $2.7\%$. Because the grid adapts to each diagram, the surfaces of two complexes are not discretized on a common grid; the node measures remain comparable because each is normalized on its own critical points.

\para{Persistence simplification.} \Cref{tab:persistence} reports the persistence thresholds for each dataset, expressed as a percentage of the instance's scalar range, along with the average number of critical points and regions before and after simplification. Persistence simplification~\cite{edelsbrunner2002topological} removes low-persistence features by cancelling saddle--extremum pairs, yielding a coarser MS complex that preserves prominent structures. \Cref{fig:persistence_curves} shows the persistence curves underlying these choices: for each instance, the curves plot the number of persistence pairs surviving at a relative threshold $p$, expressed as a percentage of the instance's scalar range, consistent with the percentages in the table. Thresholds are set just past the elbow of the curve where one exists, the point at which the number of surviving features stops decreasing rapidly~\cite{lukasczyk2020localized}. Ionization Front, Heated Cylinder, and Viscous Finger lose most of their pairs within a few percent of the range, and their thresholds sit past that drop. TOSCA drops steeply below 1\%, from 93 to 43 critical points per mesh, and is nearly flat beyond it, so 1\% sits past its elbow as well. Vortex Street has no low-persistence layer: its 157 curves nearly coincide, persistence is spread evenly up to about a quarter of the range, where only the global pair remains, and the 1\% threshold removes only the smallest pairs, about 14 of 72 per frame, leaving the 117 critical points of \cref{tab:persistence}. The curves count min--saddle and saddle--max pairs, but not the saddle--saddle pairs of the 3D Viscous Finger data. Consequently, the plotted pair counts do not directly reflect its 467 surviving critical points, most of them saddles, or the 1 to 17 regions reported in~\cref{sec:viscous} of the main paper. We did not sweep the thresholds; the main paper lists stability across thresholds as future work.

\begin{table}[t]
\centering
\caption{Persistence simplification thresholds and their effect on MS complex size. Threshold is a percentage of the scalar range per instance. Avg CPs and Avg Regions report the mean count per instance before and after simplification.}
\label{tab:persistence}
\small
\setlength{\tabcolsep}{3pt}
\begin{tabular}{@{}lc rr rr@{}}
\toprule
 & & \multicolumn{2}{c}{Before simpl.} & \multicolumn{2}{c}{After simpl.} \\
\cmidrule(lr){3-4} \cmidrule(lr){5-6}
Dataset & Thresh. & CPs & Reg. & CPs & Reg. \\
\midrule
Vortex Street    & 1\%    & 142 & 102 & 117 & 88  \\
Ionization Front & 7\%    & 516 & 270 & 34  & 29  \\
Heated Cylinder  & 11.1\%  & 282 & 164 & 45  & 30  \\
TOSCA            & 1\%    & 93  & 62  & 43  & 21  \\
Viscous Finger   & 6\%    & 656 & 105 & 467 & 10  \\
\bottomrule
\end{tabular}
\end{table}

\begin{figure*}[t]
\centering
\includegraphics[width=0.8\textwidth]{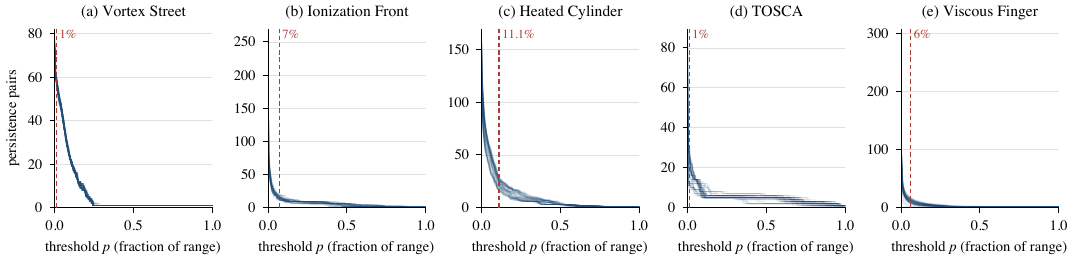}
\caption{\textbf{Persistence curves for the five datasets.} Number of surviving persistence pairs (min--saddle and saddle--max) as a function of the relative threshold $p$ (the threshold divided by the instance's scalar range); one curve per instance. Dashed lines mark the thresholds of \cref{tab:persistence}. The three datasets with a low-persistence layer, (b), (c), and (e), have their thresholds past the drop; (a) has no such layer and (d) drops below 1\%, as discussed in the text.}
\label{fig:persistence_curves}
\end{figure*}

\para{Baselines.} For the three 2D time-varying datasets, baselines are reported under uniform node weights ($\mu_i = 1/n$) and under the persistence-based measure of \Cref{sec:measures} of the main paper, that is, the persistence-image weights, which we call persistence-based below; the main-text figures show the former, while \cref{sec:appendix_timevarying} reports the latter. On the time-varying and Viscous Finger datasets, the WD cost matrix is the $\ell_2$ distance between critical point coordinates, normalized by the diagonal of the bounding box (402.0 for Vortex Street, 649.3 for Ionization Front, 3.162 for Heated Cylinder, and per-instance for Viscous Finger); on TOSCA, WD and FGW use the scalar AGD cost of~\cref{eq:scalar_cost} of the main paper. For GWD, intra-space distances are shortest-path distances on the Euclidean-weighted 1-skeleton, normalized to $[0,1]$ by dividing by the maximum value; for disconnected components, infinite distances are replaced by twice the maximum finite distance. For FGW, the feature and structure costs follow WD and GWD, with the trade-off parameter set to 0.5 on the time-varying and Viscous Finger datasets and selected by the protocol of~\cref{tab:tosca} of the main paper on TOSCA. No baseline uses entropic regularization: WD is solved as a single exact linear program, and GWD and FGW are solved by conditional gradient with an exact linear transport problem at each step.

\FloatBarrier

\section{TOSCA: Additional Results}
\label{sec:appendix_tosca}

This section adds to the classification result of~\cref{sec:tosca} of the main paper: where the errors fall, what the full distance matrices look like, and what the region coupling $\xi$ returns on pairs of shapes within and across classes.

\para{Confusion.} At $k{=}1$ leave-one-out, six of nine categories achieve perfect recall (cat, centaur, gorilla, horse, Victoria, wolf). The errors concentrate between David and Michael: 3 of 7 David meshes and 5 of 20 Michael meshes are misclassified, together eight of the ten errors on the dataset (\cref{tab:tosca} of the main paper). Both are male humanoid shapes with similar limb structure, so their MS complexes under the AGD field are nearly the same, which is why they are confused.

\para{Pairwise distance matrices.} \Cref{fig:tosca_distmat} shows the pairwise distance matrices for all 80 meshes under WD, GWD, FGW, and MS-COOT, with rows and columns ordered by category. Dark diagonal blocks show that within-class distances are smaller than between-class distances for every method, and the David-Michael block has low between-class distance under all four, consistent with the confusion above. On the other hand, GWD shows many low-distance off-diagonal blocks than other methods, indicating its lower discrepancy across meshes with different labels.

\begin{figure}[t]
  \centering
  \includegraphics[width=\columnwidth]{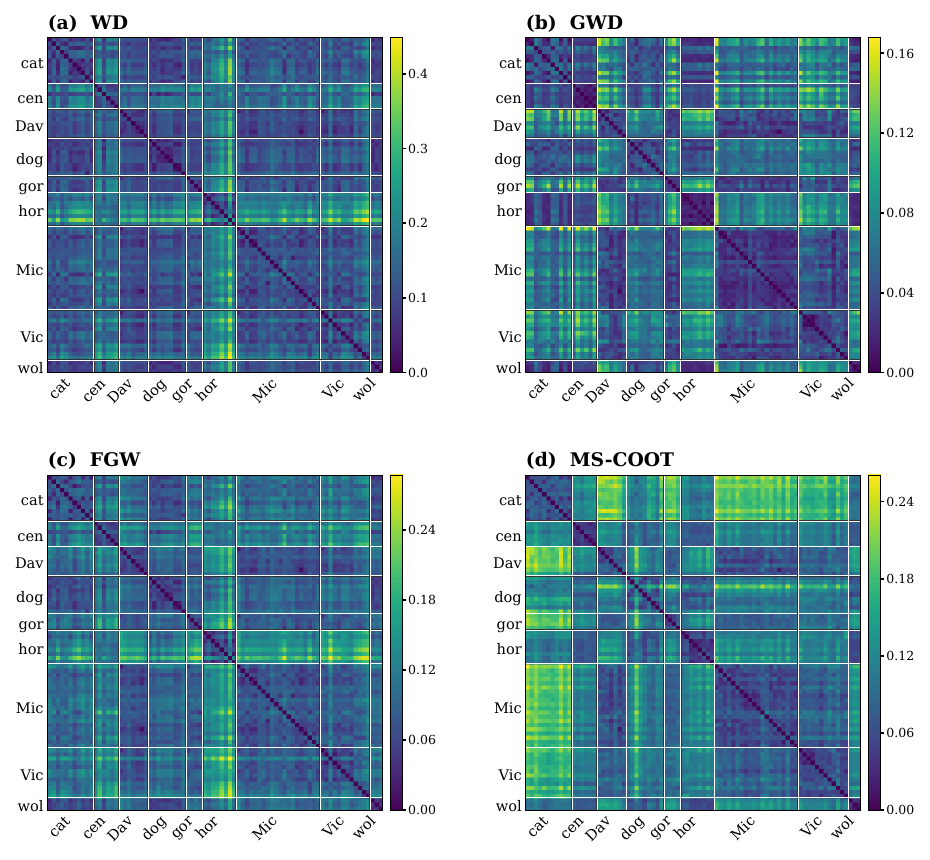}
  \caption{\textbf{Pairwise distance matrices on TOSCA} for (a)~WD, (b)~GWD, (c)~FGW, and (d)~MS-COOT. Rows and columns are ordered by category: cat~(11), centaur~(6), David~(7), dog~(9), gorilla~(4), horse~(8), Michael~(20), Victoria~(12), and wolf~(3); white lines separate category blocks. Dark diagonal blocks indicate low within-class distances. MS-COOT shows clear separation between categories with fewer low-distance off-diagonal regions, consistent with its classification recall (\cref{tab:tosca} of the main paper). The relatively low distance between the David and Michael blocks across all methods reflects their similar MS topology.}
  \label{fig:tosca_distmat}
\end{figure}

\para{Region correspondence.} \Cref{fig:tosca_intra,fig:tosca_inter} show the region coupling $\xi$ on pairs of shapes within and across classes; the source mesh is colored by its MS regions and the target mesh inherits the colors through $\xi$. Within a class, the correspondences follow the spatial organization of the regions: the upper and lower body of the centaur and the torso and limbs of the horse are aligned across poses, with the differences confined to small or boundary regions. Across classes they align only where the organization is shared: David and Michael align at the level of upper and lower body, which is the correspondence behind their confusion.
The cat and dog pair aligns at the head and limbs and is ambiguous elsewhere.

\begin{figure}[t]
  \centering
  \includegraphics[width=\columnwidth]{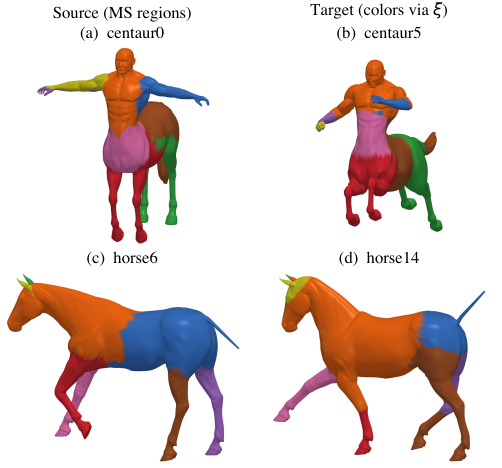}
  \caption{\textbf{MS-COOT intra-class region correspondence on TOSCA.} Left: source shapes colored by MS regions. Right: target shapes colored via transfer through the region coupling~$\xi$. (a,b)~Centaur shapes: MS-COOT preserves the relative spatial arrangement of regions (e.g., upper body vs.\ lower body), resulting in consistent correspondence of major parts despite pose variation. (c,d)~Horse shapes: limb and body regions are consistently aligned across poses, with colors preserved along corresponding legs and torso regions; minor differences reflect variations in the underlying segmentation.}
  \label{fig:tosca_intra}
\end{figure}

\begin{figure}[t]
  \centering
  \includegraphics[width=\columnwidth]{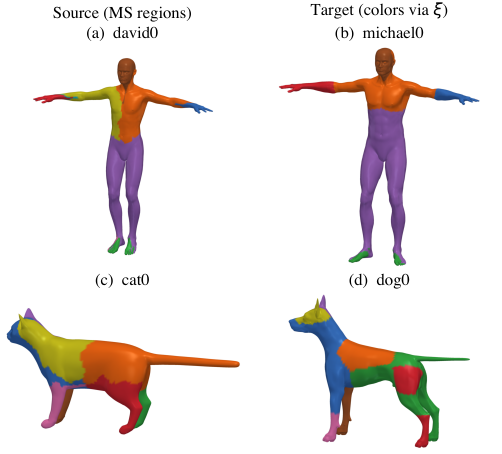}
  \caption{\textbf{MS-COOT inter-class region correspondence on TOSCA.} Left: source shapes colored by MS regions. Right: target shapes colored via transfer through the region coupling~$\xi$. (a,b)~David $\rightarrow$ Michael: regions align at a coarse level (upper vs.\ lower body), with consistent mapping of legs and partial alignment of arms. (c,d)~Cat $\rightarrow$ Dog: correspondences capture partial similarity in head and limb regions, while other regions exhibit less consistent alignment due to morphological differences.}
  \label{fig:tosca_inter}
\end{figure}

\FloatBarrier

\section{Node Measure Comparison}
\label{sec:appendix_measures}

\Cref{sec:measures} motivates the persistence-image (PI) node
measure, with controlled examples in \cref{sec:controlled}
illustrating its behavior.
Here, we further compare uniform, raw-persistence, and
PI-based measures on the clean/noisy sinusoidal pair
of \cref{fig:sinusoidal}, evaluating correspondence accuracy
against spatial ground truth.
All other configurations are held at their default values except $\alpha=0.001$.

Because the two fields share a grid, reference region
correspondences are determined by pixel-set overlap, and
critical-point (CP) correspondences by mutual nearest
neighbors of the same type.
The noise introduces 16 additional CPs and 11 regions,
primarily associated with low-persistence features.
Five clean regions are split by the perturbation, identified
when a second noisy region overlaps at least 25\% of
the clean region's pixels.
We exclude these five regions from the region-accuracy
denominator, leaving 31 unsplit regions, and evaluate
split recovery separately.

For the main comparison, we select the lowest-objective
solution from the deterministic initialization of
\cref{sec:appendix_implementation} and 40 random
initializations.
Repeating this procedure with five random seeds yields
identical correspondence accuracies and error sets.

\begin{table}[t]
\centering
\caption{Node-measure comparison on the clean/noisy sinusoidal
pair. Region and CP accuracy count reference correspondences
recovered by $\arg\max$ of the respective couplings.
Mass is the fraction of $\xi$ assigned to reference region
pairs, including split correspondences.
Split recovery counts cases whose two highest-mass targets
are the correct noisy-region pieces.
Displacement is the median distance between matched region
centers, in units of the $100\times100$ domain.
The ground-truth row gives reference values, not an
optimized transport coupling.}
\label{tab:measures}
\small
\setlength{\tabcolsep}{3.5pt}
\begin{tabular}{@{}lccccc@{}}
\toprule
$\mu$ & Region & Mass & Split & CP & Displ. \\
\midrule
Uniform           & 28/31 & 0.610 & 4/5 & 43/49 & 1.33 \\
Raw persistence   & 6/31  & 0.149 & 0/5 & 7/49  & 45.51 \\
Persistence image & \textbf{30/31} & \textbf{0.652}
                  & \textbf{4/5} & \textbf{44/49}
                  & \textbf{1.23} \\
\midrule
Ground truth      & 31/31 & 1.000 & 5/5 & 49/49 & 1.07 \\
\bottomrule
\end{tabular}
\end{table}

\Cref{tab:measures} reports the quantitative results, and
\cref{fig:measures} visualizes the correspondences.
PI-based weighting achieves the highest region and CP
accuracies, recovering 30/31 regions and 44/49 CPs,
compared with 28/31 and 43/49 under uniform weighting.
Both recover four of the five noise-induced splits.
Their incorrect region assignments are confined to
immediate neighbors, with the single PI-based error also
appearing among the three uniform-weight errors.
The median region-center displacements are 1.23 and
1.33 domain units, respectively, both below 1\% of
the domain diagonal.

In contrast, raw-persistence weighting recovers only
6/31 regions and 7/49 CPs, with no correctly recovered
splits.
Its correspondence follows an approximately
diagonal-reflected spatial orientation rather than the
alignment of the two fields, producing a median
region-center displacement of 45.51 units
(32\% of the domain diagonal).

\begin{figure}[t]
\centering
\includegraphics[width=0.9\columnwidth]
{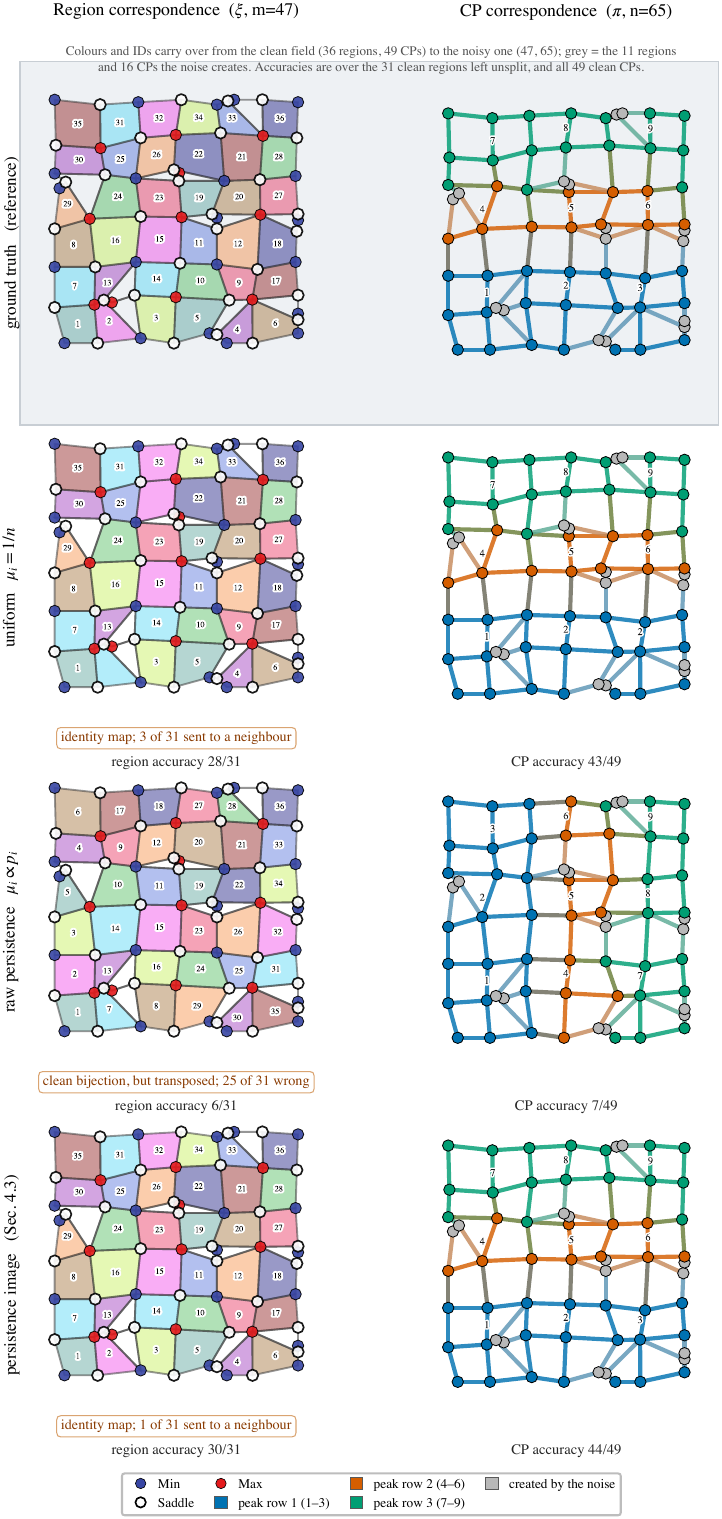}
\caption{\textbf{Correspondences under three node measures.}
Rows show the ground-truth reference, uniform weighting,
raw-persistence weighting, and PI-based weighting.
Left: region correspondence from $\xi$, with noisy-field
regions colored and indexed according to their matched
clean-field regions.
Right: CP correspondence from $\pi$, with maxima colored
by their row bands in the clean field and indexed 1--9;
noise-created CPs are gray.
Uniform and PI-based weighting largely preserve spatial
correspondence, whereas raw persistence produces an
approximately diagonal-reflected orientation.
The reported method results use the lowest-objective
solutions from the standard multistart procedure.
The clean reference is shown in~\cref{fig:sinusoidal}(e,g)
of the main paper.}
\label{fig:measures}
\end{figure}

\para{Objective preference across orientations.}
To examine the poor correspondence under raw-persistence
weighting, we compare converged solutions exhibiting
spatially aligned and diagonally reflected correspondences.
The aligned solutions are obtained from the standard
initializations, while deliberately reflection-oriented
initial couplings are additionally used to obtain reflected
solutions for comparison.
All reported objectives are evaluated at converged,
feasible transport couplings, rather than at the initial
couplings or discrete correspondence matrices.

As shown in \cref{tab:candidates}, uniform and PI-based
weighting assign lower objectives to their spatially
aligned solutions.
Raw persistence instead assigns a 38\% lower objective
to the reflected solution.
Thus, among the solutions obtained, the raw-persistence
objective favors an orientation that is inconsistent with
the spatial reference.

\begin{table}[t]
\centering
\caption{Objective values of converged MS-COOT solutions
with spatially aligned and diagonally reflected
correspondences.
The lower objective within each measure is bold.
Reflected solutions are also obtained using deliberately
reflection-oriented initializations.}
\label{tab:candidates}
\small
\setlength{\tabcolsep}{4pt}
\begin{tabular}{@{}lccc@{}}
\toprule
$\mu$ & Aligned & Reflected & Lower objective \\
\midrule
Uniform
& \textbf{0.005396} & 0.005487 & Aligned \\
Raw persistence
& 0.005789 & \textbf{0.003595} & Reflected \\
Persistence image
& \textbf{0.004444} & 0.004633 & Aligned \\
\bottomrule
\end{tabular}
\end{table}

\para{Effect of node-weight variation.}
The poor correspondence under raw-persistence weighting
cannot be explained by insufficient suppression of
noise-created CPs alone.
Raw persistence assigns only 8.9\% of the noisy field's
node mass to these CPs, compared with 19.2\% under
PI-based weighting, yet yields substantially worse
correspondences.

The two measures also differ in how their node weights
change under perturbation.
Across reference CP correspondences, the median relative
weight change is 20\% under raw persistence, compared
with 6\% under PI-based weighting.
This greater variation is consistent with the poorer
spatial alignment under raw persistence and the
persistence-pairing sensitivity illustrated in
\cref{fig:twopeak} of the main paper.
Because the node measure also determines the region
marginals, these changes affect both $\pi$ and $\xi$.
The results suggest that preserving weight consistency
across corresponding features can be more important
for correspondence recovery than suppressing
low-persistence features alone.

\para{Initialization sensitivity.}
Raw persistence is the only measure whose correspondence
accuracy changes under the standard initialization
procedure.
The deterministic initialization yields 26/31 correct
region correspondences, whereas random starts also reach
the diagonally reflected solution, which has a lower
objective but recovers only 6/31 regions.
Uniform and PI-based weighting yield identical accuracies
and error sets under the standard starts; their reflected
solutions in \cref{tab:candidates} are obtained using
separate, deliberately designed initializations.
We report the lowest-objective solutions from the
standard multistart procedure consistently across measures.

Overall, PI-based weighting provides the most accurate
correspondences among the three measures on this example.
The comparison supports its use as a practical default,
while illustrating that emphasizing persistence alone
does not guarantee spatially meaningful transport.

\FloatBarrier

\section{Time-Varying Datasets: Additional Results}
\label{sec:appendix_timevarying}

This section reports every configuration of the comparison in~\cref{sec:results} of the main paper on the three time-varying datasets, Vortex Street, Ionization Front, and Heated Cylinder: the three baselines under uniform and persistence-based node weights, and MS-COOT under uniform and persistence-based weights. 
Vortex Street and Ionization Front are analyzed in~\cref{sec:vortex,sec:ionization} of the main paper, so the section begins with the Heated Cylinder experiment, which is reported here in full. It then gives the adjacent-step distance curves for Ionization Front, the analysis of the off-diagonal structure in its MS-COOT matrix, and finally the full distance matrices of all eight configurations on each dataset (\cref{fig:supp_grid_vs,fig:supp_grid_if,fig:supp_grid_hc}).

\para{Heated Cylinder.}
The Heated Cylinder dataset is a 2D flow simulation around a
heated obstacle~\cite{cgl,gerrisflowsolver,GuntherGrossTheisel2017},
containing 2{,}001 timesteps.
We analyze timesteps 800--899, indexed 0--99 below.
After persistence simplification at 11.1\% of the scalar
range, each timestep contains 31--57 critical points
(mean 45) and 16--43 regions (mean 30).
The sequence exhibits a marked increase in region count
at timestep $56{\to}57$, providing an opportunity to examine how the methods respond to changes in the region decomposition.

\begin{figure*}[t]
\centering
\includegraphics[width=0.9\textwidth]
{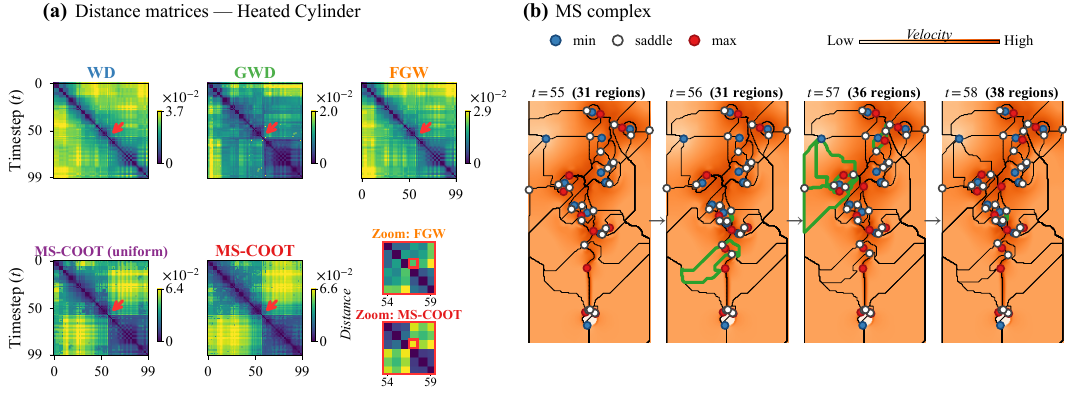}
\vspace{-2mm}
\caption{\textbf{Heated Cylinder:
structural transition at timestep $56{\to}57$.}
(a) Pairwise distance matrices for WD, GWD, and FGW
with uniform node weights, and MS-COOT with uniform
and persistence-image (PI) weights.
All methods register the transition (red arrows),
while MS-COOT exhibits a clearer separation between
timesteps before and after the transition.
Insets show timesteps 54--59 for FGW and MS-COOT.
(b) MS complexes overlaid on the velocity magnitude
field at timesteps 55--58.
Green outlines highlight changes in the separatrices.
Between timesteps 56 and 57, additional separatrices
subdivide several regions, increasing the region count
from 31 to 36.}
\vspace{-3mm}
\label{fig:heated_cylinder}
\end{figure*}

\Cref{fig:heated_cylinder}(a) shows that all methods
register the transition near timestep $56{\to}57$,
but differ in their separation of the sequence.
MS-COOT exhibits two blocks with relatively low
within-block and high between-block distances,
whereas the baselines show less pronounced separation.
We quantify this behavior in \cref{tab:hc_separation}
using the silhouette coefficient of the two-block
labelling ($t\leq56$ and $t\geq57$), the ratio
of mean between-block to mean within-block distance
(B/W), and the descending rank of $d(56,57)$
among all 99 adjacent-step distances.

\begin{table}[t]
\centering
\caption{Heated Cylinder structural separation under
eight method--measure configurations.
Silhouette evaluates the two-block labelling
$t\leq56$ versus $t\geq57$ using the pairwise
distance matrix.
B/W is the ratio of mean between-block to mean
within-block distance.
Rank is the descending position of $d(56,57)$
among the 99 adjacent-step distances
(1 denotes the largest).}
\label{tab:hc_separation}
\small
\setlength{\tabcolsep}{4pt}
\begin{tabular}{@{}llccc@{}}
\toprule
Method & Weights & Silhouette & B/W & Rank \\
\midrule
WD      & Uniform & 0.286 & 1.39 & 3 \\
GWD     & Uniform & 0.395 & 1.64 & 4 \\
FGW     & Uniform & 0.318 & 1.44 & 2 \\
WD      & PI      & 0.337 & 1.51 & 3 \\
GWD     & PI      & 0.456 & 1.80 & 1 \\
FGW     & PI      & 0.373 & 1.57 & 3 \\
MS-COOT & Uniform & 0.532 & 2.13 & 1 \\
MS-COOT & PI      & \textbf{0.594}
                   & \textbf{2.47} & 1 \\
\bottomrule
\end{tabular}
\end{table}

PI-weighted MS-COOT achieves the strongest separation,
with a silhouette of 0.594 and a B/W ratio of 2.47.
Uniform-weight MS-COOT also exceeds every baseline,
including those using PI weights
(silhouette 0.532 versus 0.456 for the strongest baseline).
Both MS-COOT variants rank $56{\to}57$ as the largest
adjacent-step distance, while all baseline configurations
rank it among their four largest.
Thus, the transition is detected by all methods, but
MS-COOT more distinctly separates the two groups
of timesteps.

\para{Relation to region decomposition.}
The MS complexes in \cref{fig:heated_cylinder}(b)
provide a structural explanation for the transition.
From timestep 55 to 56, the region count remains at 31,
whereas at $56{\to}57$ it increases to 36 as additional
separatrices subdivide several existing regions.
The count reaches 38 at timestep 58.
The full region-count trajectory
(\cref{fig:HC_region_count}) shows that the number
of regions remains at a generally higher level
after the transition.
Because these counts are obtained directly from the
extracted MS complexes, independently of the distance
computations, they provide complementary evidence
of a change in the underlying topological decomposition.

\begin{figure}[t]
\centering
\includegraphics[width=0.95\columnwidth]
{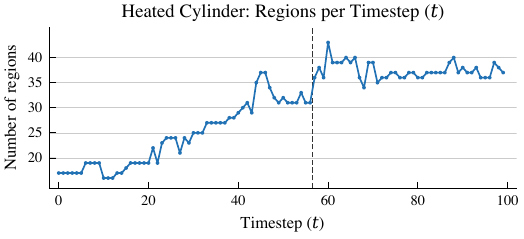}
\vspace{-2mm}
\caption{\textbf{Region-count evolution in Heated Cylinder.}
The number of regions is computed directly from the
MS complexes, independently of the pairwise distances.
A marked increase occurs at timestep $56{\to}57$
(dashed line), followed by a generally higher region
count over the remaining sequence.}
\label{fig:HC_region_count}
\vspace{-2mm}
\end{figure}

The additional separatrices alter both the region
partition and its relationships to critical points.
In MS-COOT, these changes are represented through
the critical-point--region relational function $\omega$
and the jointly optimized couplings $\pi$ and $\xi$.
Consequently, changes in region subdivision can
contribute to the distance beyond changes in the
critical-point configuration alone.
The stronger separation under uniform-weight MS-COOT
indicates that persistence weighting is not necessary
for this structural transition to stand out.
PI-based weighting further increases the separation,
showing that feature significance can modulate the
response to the same structural changes.

These observations are consistent with the controlled
examples in \cref{sec:controlled} of the main paper:
MS-COOT responds to changes in region geometry and
separatrices under uniform weights, while PI-based
weights additionally account for feature significance.
Heated Cylinder also complements Ionization Front
(\Cref{sec:ionization} of the main paper), where the highlighted
region merge becomes the largest adjacent-step change
only under PI-based weighting.
Here, the region subdivisions already produce a
prominent response with uniform weights, which
PI-based weighting further strengthens.
Together, the two datasets illustrate that the
contribution of persistence weighting depends on
the structural changes being compared.

\para{Adjacent-step distances.}
\Cref{fig:supp_consecutive} shows the adjacent-step distances
$d(t,t{+}1)$ for all 29 consecutive pairs of Ionization Front
under five configurations: WD, GWD, and FGW with uniform
weights, and MS-COOT with uniform and persistence-image
(PI) weights.
Each sequence is divided by its mean adjacent-step distance
to facilitate comparison across methods.
The region merge at $3{\to}4$, discussed in
\cref{sec:ionization} of the main paper, produces the
largest adjacent-step distance under PI-weighted MS-COOT,
but not under any baseline or uniform-weight MS-COOT,
where it ranks 21st.
The complete profile places this highlighted transition
in the context of other temporal changes.
The corresponding Vortex Street profiles are already
reported in \cref{fig:vortexstreet}(b) of the main paper.

\begin{figure}[t]
\centering
\includegraphics[width=\columnwidth]
{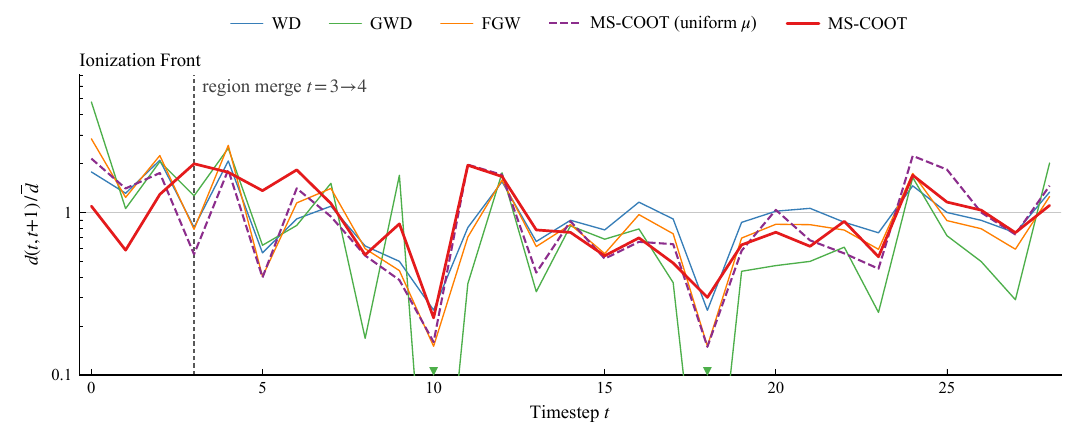}
\caption{\textbf{Ionization Front adjacent-step distances.}
Distances $d(t,t{+}1)$ for all 29 consecutive pairs under
five configurations, each normalized by its mean over
the sequence and displayed on a logarithmic scale.
The highlighted $3{\to}4$ region merge is the largest
step under PI-weighted MS-COOT, but not under the other
configurations.
Values below 0.1, all from GWD, are displayed as
markers at the plotting floor.}
\label{fig:supp_consecutive}
\end{figure}

\para{Off-diagonal distance variations.}
The Ionization Front distance matrices
(\cref{fig:supp_grid_if}) exhibit different degrees
of local variation across methods.
WD and FGW show relatively smooth increases in
distance with temporal separation, whereas GWD
exhibits more irregular blockwise patterns.
MS-COOT retains a comparatively smooth structure
near the diagonal but contains some localized
variations for temporally distant frame pairs,
particularly toward the upper-right and lower-left
portions of the matrices.

One possible explanation is ambiguity in structural
correspondence when the compared MS complexes
differ substantially.
The default MS-COOT sample cost distinguishes
critical-point types but does not directly constrain
their spatial correspondence.
Although the hypernetwork function encodes
geometric relationships between critical points
and regions, substantially different complexes
may admit multiple plausible structural alignments,
potentially affecting the resulting distances.
The correspondence ambiguity observed in the
controlled examples and the node-measure comparison
provides related evidence, although its contribution
to these temporal variations has not been established.

Another possible contributor is the per-complex
normalization of $\omega$ by its maximum entry,
which makes the distribution of normalized
relational values sensitive to unusually long
shortest paths.
These effects may contribute to the observed
variations, but their relative importance remains
unclear.
A systematic analysis of correspondence ambiguity
and alternative relational normalizations is left
for future work.

\para{Additional distance matrices.}
\Cref{fig:supp_grid_vs,fig:supp_grid_if,fig:supp_grid_hc}
provide pairwise distance matrices for the eight
method--measure configurations: WD, GWD, FGW, and
MS-COOT, each with uniform and PI-based weights.
These additional comparisons complement the
representative configurations shown in the main paper
and illustrate how persistence weighting affects
the temporal distance patterns across methods.
For Vortex Street, the off-diagonal bands indicate
recurrence after approximately 74--75 frames,
consistent with the periodicity reported in~\cref{sec:vortex}.
For Ionization Front, the matrices show the evolving
dissimilarity as the front advances, together with
the off-diagonal variations discussed above.
For Heated Cylinder, both MS-COOT weighting schemes
show a more pronounced separation around the
structural transition at timestep $56{\to}57$
than the evaluated baselines
(\cref{tab:hc_separation}).

\begin{figure*}[p]
\centering
\includegraphics[width=0.92\textwidth]
{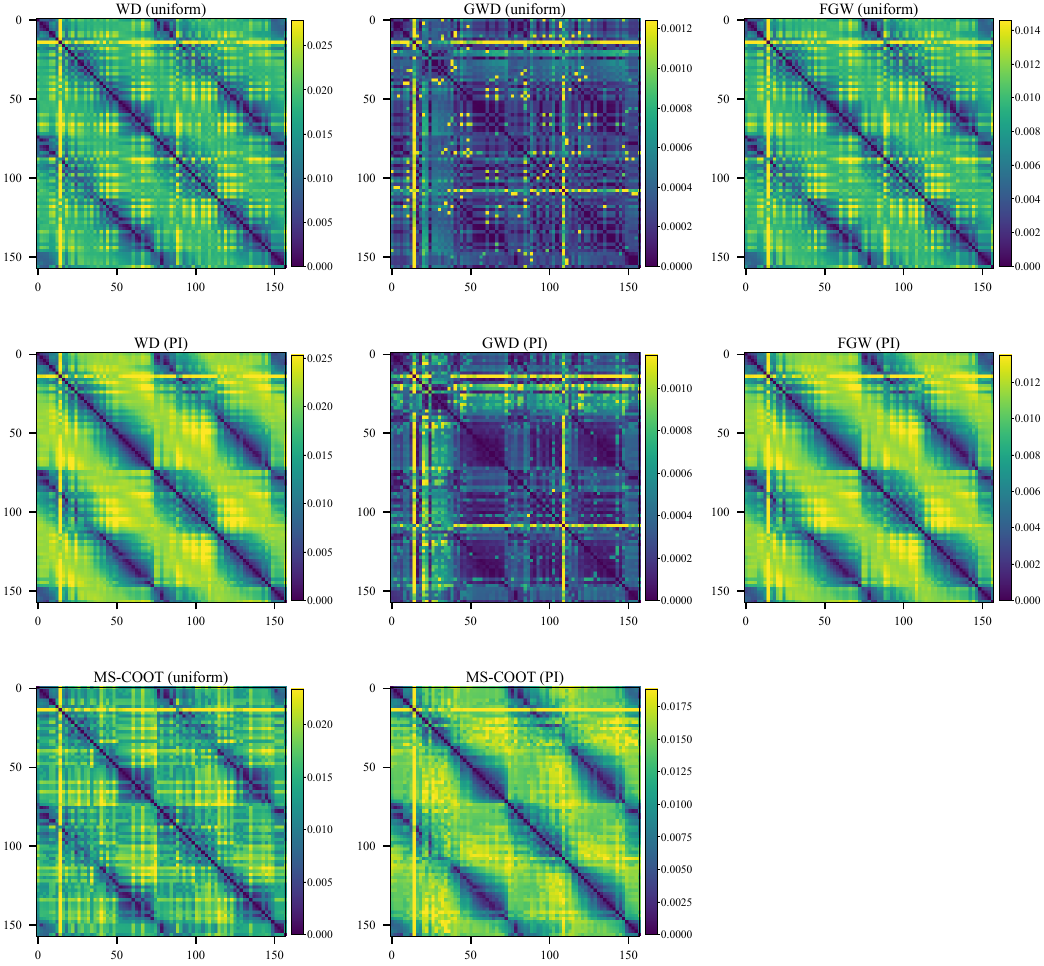}
\caption{\textbf{Vortex Street distance matrices under
eight configurations.}
WD, GWD, FGW, and MS-COOT are compared using
uniform and PI-based node weights.
Exhibited matrices use the stride-2 sequence
$t=0,2,\ldots,156$.
\Cref{fig:vortexstreet}(a) of the main paper
shows five representative configurations on
all 157 timesteps.
The off-diagonal bands indicate a recurrence period
of approximately 74--75 frames, with differences
in band prominence across methods and measures.}
\label{fig:supp_grid_vs}
\end{figure*}

\begin{figure*}[p]
\centering
\includegraphics[width=0.92\textwidth]{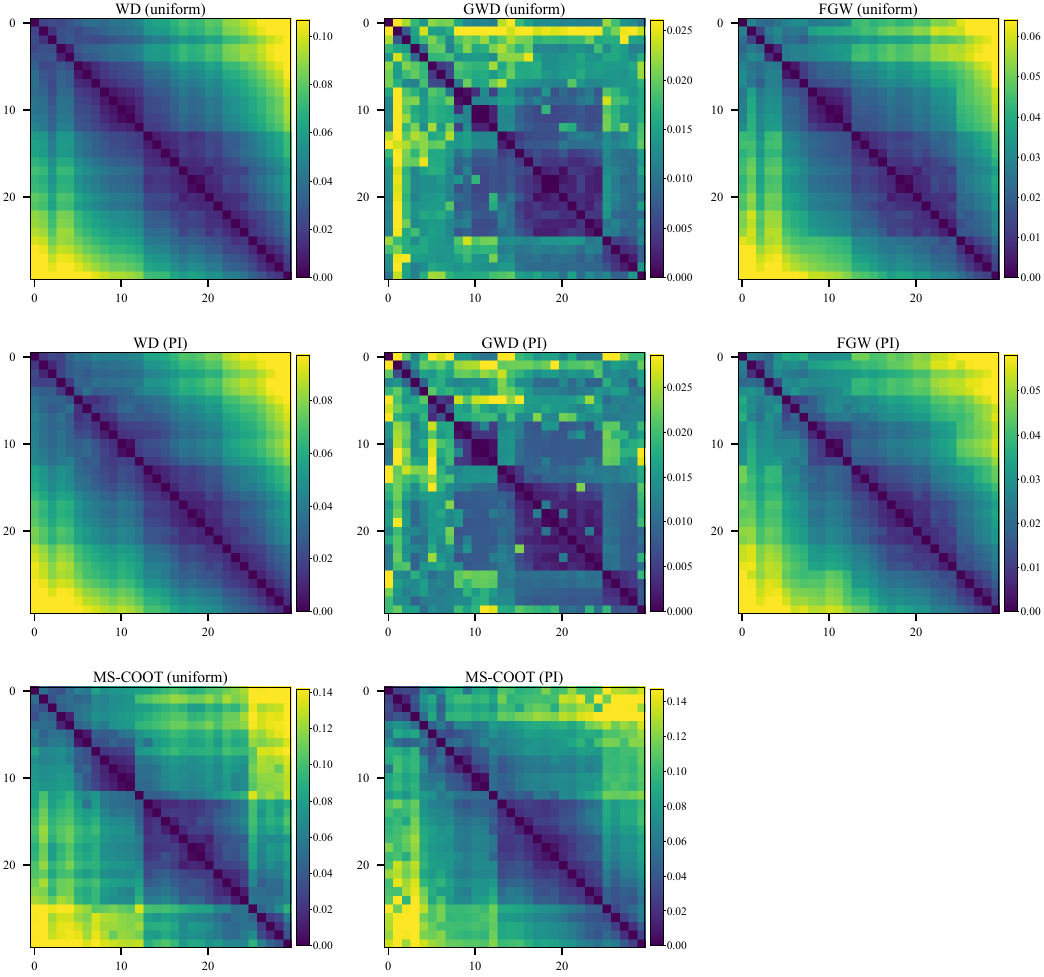}
\caption{\textbf{Ionization Front, all eight configurations.} Same layout as \cref{fig:supp_grid_vs}. The $3{\to}4$ transition of~\cref{fig:ionfront} of the main paper is the largest adjacent step for MS-COOT under persistence-based weights and for no other configuration plotted in \cref{fig:supp_consecutive}.}
\label{fig:supp_grid_if}
\end{figure*}

\begin{figure*}[p]
\centering
\includegraphics[width=0.92\textwidth]{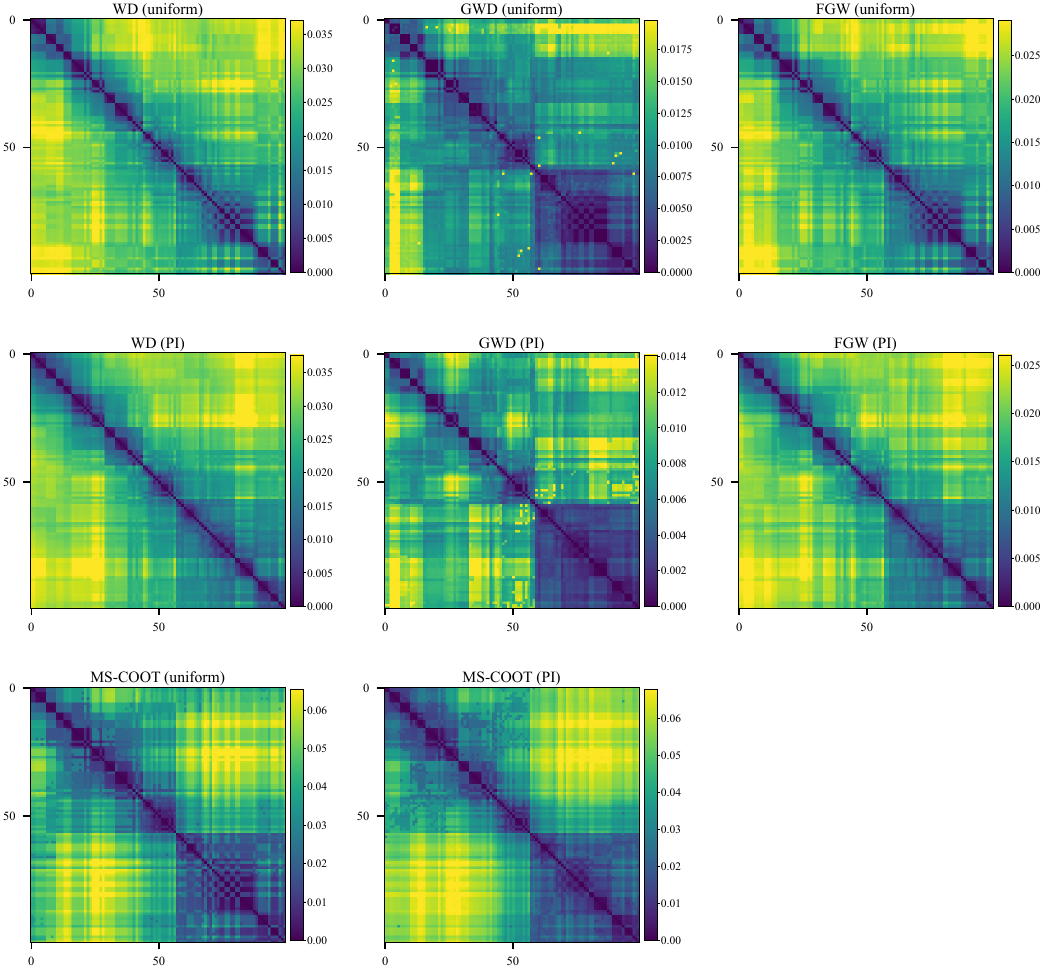}
\caption{\textbf{Heated Cylinder, all eight configurations.} Same layout as \cref{fig:supp_grid_vs}. The regime boundary at $56{\to}57$ of \cref{fig:heated_cylinder} appears in every configuration; the two-block structure is sharpest for MS-COOT under either weighting.}
\label{fig:supp_grid_hc}
\end{figure*}

\FloatBarrier

\section{Consistency of the Two Couplings}
\label{sec:appendix_consistency}

MS-COOT jointly optimizes the critical-point coupling $\pi$ and region coupling $\xi$ through the structural term of~\cref{eq:mscoot} in the main paper, but does not explicitly enforce consistency between their correspondences.
In particular, a pair of regions receiving high transport mass in $\xi$ need not have their boundary critical points matched to one another by $\pi$.
We evaluate how frequently such disagreements occur and how they relate to the confidence of region correspondences.

\para{Boundary consistency.}
For a source region $R_k^f$, let $l^\ast=\arg\max_l\xi[k,l]$ denote its highest-mass target region, and let $\partial R$ denote the set of critical points on a region's boundary.
We define the boundary consistency of $R_k^f$ as
\[
c_k =
\frac{
\sum_{i\in\partial R_k^f}
\sum_{j\in\partial R_{l^\ast}^g}
\pi[i,j]
}{
\sum_{i\in\partial R_k^f}\mu_i^f
}.
\]
This measures the fraction of transport mass from the source region's boundary critical points that reaches the boundary of its highest-mass target.
A value of $c_k=1$ indicates that all such mass reaches the target boundary, while $c_k=0$ indicates
 hat none does.

For descriptive purposes, we define a boundary conflict as $c_k<0.5$ and a severe conflict as $c_k<0.25$.
These thresholds quantify disagreement between the two couplings, rather than incorrect spatial correspondence.
In particular, $c_k$ considers only the highest-mass target region and may be low when transport is distributed across multiple regions, as can occur under region splitting.
We evaluate the measure in both transport directions.

\para{Overall consistency.}
\Cref{tab:consistency} reports the results for 29 adjacent pairs of Ionization Front and 99 adjacent pairs of Heated Cylinder, using the default MS-COOT configuration ($\alpha=0.5$, $\varepsilon=0.001$, and persistence-image weights).
The median boundary consistency is 0.751 on Ionization Front and 0.968 on Heated Cylinder.
Boundary conflicts occur for 28.1\% and 7.8\% of evaluated region instances, respectively, while severe conflicts account for 8.4\% and 2.7\%.
Thus, boundary transport is largely consistent between the two couplings, particularly on Heated Cylinder, although disagreement is not negligible.

\begin{table}[t]
\centering
\caption{Boundary consistency on adjacent timestep pairs of Ionization Front (29 pairs) and Heated Cylinder (99 pairs), with both transport directions pooled. Regions counts evaluated source-region instances. Conflict denotes $c_k<0.5$, and severe conflict denotes $c_k<0.25$.}
\label{tab:consistency}
\small
\setlength{\tabcolsep}{4pt}
\begin{tabular}{@{}lrccc@{}}
\toprule
Dataset & Regions & Med.\ $c_k$ & Conflict & Severe \\
\midrule
Ionization Front & 1{,}528 & 0.751 & 28.1\% & 8.4\% \\
Heated Cylinder  & 5{,}940 & 0.968 & 7.8\% & 2.7\% \\
\bottomrule
\end{tabular}
\end{table}

\para{Relation to region-match confidence.}
To examine where disagreements occur, we define the confidence of the region match as
\[
q_k = \frac{\xi[k,l^\ast]}{\nu_k^f},
\]
the fraction of the source region's mass assigned to its highest-mass target.
\Cref{tab:consistency_conf} reports boundary conflict rates across confidence intervals.

Conflicts are concentrated among diffuse region matches.
When $q_k\geq0.75$, the conflict rate is only 2.2\% on Ionization Front and 0.9\% on Heated Cylinder.
For $q_k<0.5$, these rates increase to approximately 35\% and 17\%, respectively.
The conflict rate decreases across successive confidence intervals on both datasets.
Thus, concentrated region transport is generally associated with stronger agreement between $\xi$ and the boundary CP transport in $\pi$.

\begin{table}[t]
\centering
\caption{Boundary conflict rate ($c_k<0.5$)
by region-match confidence $q_k$.
Region counts in each interval, Ionization Front /
Heated Cylinder, are 202/32, 951/1{,}959,
239/1{,}947, and 136/2{,}002, respectively.}
\label{tab:consistency_conf}
\small
\setlength{\tabcolsep}{6pt}
\begin{tabular}{@{}lcc@{}}
\toprule
Confidence $q_k$ & Ionization Front & Heated Cylinder \\
\midrule
$<0.25$         & 45.0\% & 62.5\% \\
$[0.25,0.50)$   & 32.3\% & 15.9\% \\
$[0.50,0.75)$   & 12.1\% & 5.8\% \\
$\geq0.75$      & 2.2\%  & 0.9\% \\
\bottomrule
\end{tabular}
\end{table}

Region size shows a less consistent relationship with boundary disagreement.
On Heated Cylinder, conflicting regions have a median size of 198 pixels, compared with 411 pixels for the remaining regions.
On Ionization Front, the corresponding medians are 329 and 320 pixels.
These results suggest that region size may contribute to disagreement in some settings, but is not a consistent predictor across datasets.

For example, at Heated Cylinder timestep $21{\to}22$, region 16 assigns 61\% of its mass to region 7, yet its boundary consistency is $c_k\approx0$.
This illustrates that even a dominant region match need not be supported by the corresponding critical-point transport.

Overall, the two couplings exhibit substantial boundary agreement, especially for concentrated region matches, but consistency is not guaranteed by the current formulation.
The region-match confidence $q_k$ can indicate where the two couplings are more likely to agree, although neither $q_k$ nor $c_k$ alone establishes spatial correspondence correctness.
For uncertain region matches, inspecting $\pi$ together with $\xi$ provides a more complete interpretation of the recovered correspondences.

\FloatBarrier

\section{Critical Point Couplings Across Methods}
\label{sec:appendix_cp_matching}

WD, GWD, and FGW produce critical-point (CP) couplings, as does MS-COOT through $\pi$.
We compare these couplings on the clean/noisy sinusoidal pair with known reference correspondences (\cref{fig:sinusoidal}) and briefly examine the cancelled CPs in the Ionization Front example (\cref{fig:ionfront}).
The goal is to distinguish CP correspondence performance from the additional region-level correspondence provided by MS-COOT.

\para{Synthetic pair with reference correspondences.}
We use the 49 reference CP correspondences of the clean/noisy sinusoidal pair established in \cref{sec:appendix_measures}.
The 16 noise-induced CPs have no reference partners and are excluded from the accuracy denominator, but remain in the target coupling and its probability marginals.
The baselines use the costs of \cref{sec:implementation} in the main paper with uniform or persistence-image (PI) marginals.
WD uses an exact linear transport solver, while GWD and FGW use conditional-gradient optimization with exact linear transport subproblems.
MS-COOT uses the couplings reported in \cref{tab:measures}, with the configuration of \cref{fig:sinusoidal} in the main paper.

\Cref{tab:cp_matching} reports three statistics.
Accuracy counts correct partners recovered by row-wise $\arg\max$.
Mass-on-truth is the fraction of source CP mass transported to the reference pairs.
Dominant rows measures the proportion of source CPs whose largest coupling entry exceeds half their row mass. 
We note that the latter quantifies coupling concentration, not correspondence correctness.

\begin{table}[t]
\centering
\caption{CP coupling comparison on the clean/noisy sinusoidal pair with 49 reference correspondences.
Accuracy counts correct row-wise $\arg\max$ matches.
Mass-on-truth is the fraction of source CP mass transported to reference pairs.
Dominant rows is the fraction of source rows whose largest entry exceeds 50\% of the row mass.
All methods are evaluated against the same reference correspondences, but mass-on-truth is constrained by the selected transport marginals.}
\label{tab:cp_matching}
\small
\setlength{\tabcolsep}{3pt}
\begin{tabular}{@{}llccc@{}}
\toprule
Method & Weights & Accuracy & Truth mass & Dominant rows \\
\midrule
WD      & Uniform & 47/49 & 0.723 & 1.000 \\
GWD     & Uniform & 0/49  & 0.005 & 1.000 \\
FGW     & Uniform & 41/49 & 0.649 & 0.980 \\
WD      & PI      & 48/49 & 0.789 & 0.980 \\
GWD     & PI      & 6/49  & 0.117 & 0.939 \\
FGW     & PI      & 48/49 & 0.768 & 0.980 \\
MS-COOT & Uniform & 43/49 & 0.585 & 0.837 \\
MS-COOT & PI      & 44/49 & 0.684 & 0.898 \\
\bottomrule
\end{tabular}
\end{table}

WD and FGW with PI weights each recover 48/49 reference
correspondences, compared with 44/49 for PI-weighted
MS-COOT.
They also place more transport mass on reference pairs
(0.789 and 0.768 versus 0.684).
These methods directly incorporate cross-field Euclidean
distances between CP locations, whereas the tested
MS-COOT configuration encodes geometry through
the within-complex CP--region relations in $\omega$.
Its sample cost penalizes type mismatches with
$\alpha=0.001$, providing relatively weak direct
attribute constraints.
Three of the five incorrect MS-COOT correspondences
involve different CP types.
Thus, under these configurations, direct spatial costs
yield more accurate CP correspondences on this pair.

The mass-on-truth scores also depend on the marginals.
Because the noisy field contains 65 CPs, uniform
marginals limit the total mass on the 49 distinct
reference pairs to $49/65\approx0.754$.
Coupling concentration likewise does not guarantee
correct correspondence: uniform-weight GWD has
a dominant target in every source row but recovers
none of the 49 reference partners.
Its poor spatial alignment is consistent with the
ambiguity of relational matching without direct
cross-field spatial information.
These observations distinguish correspondence accuracy
from transport concentration and marginal effects.

\para{Cancelled CPs in Ionization Front.}
At the region merge between timesteps $3$ and $4$ (\cref{sec:ionization},~\cref{fig:ionfront}), a saddle and a maximum disappear and have no surviving reference partners.
Balanced transport nevertheless assigns their mass to surviving CPs.
MS-COOT sends all of the saddle's mass to one surviving saddle and 92\% of the maximum's mass to one surviving maximum, whereas WD and FGW distribute some mass across CPs of different types.
These assignments reflect differences in the methods' costs and couplings.

\para{Discussion.}
On the synthetic pair, MS-COOT does not improve CP correspondence accuracy over WD or FGW under the tested configurations.
Its additional capability is the jointly optimized region coupling $\xi$, which directly represents transport between MS regions.
The CP-based baselines do not optimize a region-indexed coupling with its own marginals and structural objective.
The boundary-consistency analysis of \cref{sec:appendix_consistency} further examines the agreement between $\pi$ and $\xi$.
Together, these results clarify that the principal contribution of MS-COOT is joint CP--region comparison instead of improved CP-only matching.

\else
    \appendices
    \crefalias{section}{appendix}
    \crefname{appendix}{Appendix}{Appendices}
    \Crefname{appendix}{Appendix}{Appendices}
    \renewcommand{\section}[1]{%
        \refstepcounter{section}%
    }
    

\section{Notation}
\label{sec:appendix_notation}

\refstepcounter{table}
\label{tab:notation}

\section{Implementation Details}
\label{sec:appendix_implementation}

\refstepcounter{table}
\label{tab:params}

\refstepcounter{table}
\label{tab:persistence}

\refstepcounter{figure}
\label{fig:persistence_curves}

\section{TOSCA: Additional Results}
\label{sec:appendix_tosca}

\refstepcounter{figure}
\label{fig:tosca_distmat}

\refstepcounter{figure}
\label{fig:tosca_intra}

\refstepcounter{figure}
\label{fig:tosca_inter}

\section{Node Measure Comparison}
\label{sec:appendix_measures}

\refstepcounter{table}
\label{tab:measures}

\refstepcounter{figure}
\label{fig:measures}

\refstepcounter{table}
\label{tab:candidates}

\section{Time-Varying Datasets: Additional Results}
\label{sec:appendix_timevarying}

\refstepcounter{figure}
\label{fig:heated_cylinder}

\refstepcounter{table}
\label{tab:hc_separation}

\refstepcounter{figure}
\label{fig:HC_region_count}

\refstepcounter{figure}
\label{fig:supp_consecutive}

\refstepcounter{figure}
\label{fig:supp_grid_vs}

\refstepcounter{figure}
\label{fig:supp_grid_if}

\refstepcounter{figure}
\label{fig:supp_grid_hc}

\section{Consistency of the Two Couplings}
\label{sec:appendix_consistency}

\refstepcounter{table}
\label{tab:consistency}

\refstepcounter{table}
\label{tab:consistency_conf}

\section{Critical Point Couplings Across Methods}
\label{sec:appendix_cp_matching}

\refstepcounter{table}
\label{tab:cp_matching}

\fi

\end{document}